\documentclass{elsart}

\usepackage{amssymb,amsmath,graphicx,cite,xcolor}

\DeclareSymbolFont{bbold}{U}{bbold}{m}{n}
\DeclareSymbolFontAlphabet{\mathbbold}{bbold}

\newcommand{\RR}{\mathbb{R}}

\newcommand{\A}{\mathcal{A}}
\newcommand{\ud}{\,\mathrm{d}}

\newcommand{\diver}{\text{div}\,}

\def\dv{{\rm d}}

\def\undertilde#1{\mathord{\vtop{\ialign{##\crcr
$\hfil\displaystyle{#1}\hfil$\crcr\noalign{\kern1.5pt\nointerlineskip}
$\hfil\tilde{}\hfil$\crcr\noalign{\kern1.5pt}}}}}

\begin{document}

\begin{frontmatter}

\maketitle

\title{Asymmetry in crystal facet dynamics of homoepitaxy by a
    continuum model}

\author[a]{Jian-Guo Liu}\ead{jliu@math.duke.edu},
\author[a,b]{Jianfeng Lu}\ead{jianfeng@math.duke.edu},
\author[c]{Dionisios Margetis\corauthref{cor}}\ead{dio@math.umd.edu}
\corauth[cor]{Corresponding author. Tel.: (301) 405 5455; fax: (301) 314 0827.},
\author[d]{Jeremy L. Marzuola}\ead{marzuola@math.unc.edu}

\address[a]{Department of Mathematics and Department of Physics, Duke University, Durham, NC 27708}

\address[b]{Department of Chemistry, Duke University, Durham, North Carolina 27708}

\address[c]{Department of Mathematics, and Institute for Physical Science and Technology, and Center for Scientific Computation and Mathematical Modeling, University of Maryland, College Park, MD 20742}

\address[d]{Department of Mathematics, University of North Carolina, Chapel Hill, NC 27599}

\begin{abstract}
In the absence of external material deposition, crystal surfaces usually relax to become flat by decreasing their free energy. We study analytically an asymmetry in the relaxation of macroscopic plateaus, facets, of a periodic surface corrugation in 1+1 dimensions via a continuum model below the roughening transition temperature. The model invokes a continuum evolution law expressed by a highly degenerate parabolic partial differential equation (PDE) for surface diffusion, which is related to the nonlinear gradient flow \color{black} of a convex, singular surface free energy with a certain exponential mobility \color{black} in homoepitaxy. This evolution law is motivated both by an atomistic broken-bond model and a mesoscale model for crystal steps. By constructing an explicit solution to this PDE, we demonstrate the lack of symmetry in the evolution of  top and bottom facets in periodic surface profiles. Our explicit, analytical solution is compared to numerical simulations of the continuum law via a regularized surface free energy.  
\end{abstract}


\end{frontmatter}

\section{Introduction}
\label{sec:intro}

The epitaxial growth and relaxation of crystals include kinetic processes by  which atoms are deposited from above, and are adsorbed and diffuse on a substrate to form solid films or other nanostructures. Hence, the crystal surface undergoes morphological changes~\cite{PimpinelliVillain98,JeongWilliams99,Misbah10}. If the crystal of the film matches that of the substrate, the processes pertain to homoepitaxy. Below the roughening transition temperature, macroscopic plateaus, called {\em facets}, may form. Their evolution is linked to various nanoscale phenomena~\cite{Misbah10}; for example, the stability of semiconductor quantum dots and the wetting/dewetting of crystal surfaces~\cite{Chame2013}.

In this paper, we study implications of a continuum model based on a singular-diffusion partial differential equation (PDE)
satisfied by the height profile in crystal surface relaxation, in the
absence of external material deposition, in 1+1 dimensions. This evolution law
encompasses continuum thermodynamics and mass
conservation. The model is related to a
nonlinear, weighted $H^{-1}$ gradient flow \color{black} for a convex, singular surface free
energy in homoepitaxy. The PDE is motivated by the continuum limit of
the following models: (i) a {\em mesoscale} theory of line defects,
steps, under diffusion-limited kinetics in monotone step
trains~\cite{BCF51,DM_Kohn06}; and (ii) a family of {\em atomistic},
broken-bond models, in which the kinetic rates obey the Arrhenius law
involving the energy barriers for atom hopping~\cite{KDM,MW1}.

Physically, our continuum model reflects the presence of strong, isotropic {\em stiffness} of steps. This notion of step stiffness is related to the energy cost to create a step, and affects the local-equilibrium density, $\varrho_s$, of adsorbed atoms (adatoms). By the Gibbs-Thomson relation at equilibrium~\cite{Rowlinson,Sethna96}, \color{black} this $\varrho_s$ is an exponential function of the step chemical potential, $\mu_{\rm s}$, scaled by the Boltzmann energy, $k_B T$. The $\mu_{\rm s}$  is defined as the change per atom in the step energy; and in principle expresses the joint effect of step stiffness and step-step interactions~\cite{JeongWilliams99,Kukta02-I,Kukta02-II}. We assume that $|\mu_{\rm s}|$ may be of the same order as or larger than  $k_BT$; thus, the exponential dependence of $\varrho_s$ on $\mu_s$ cannot be neglected. A similar chemical potential was used in \cite{KDM} in the setting of adatom rates in order to derive continuum equations for the height profile from an atomistic perspective. At the continuum level, the assumption of an exponential law for $\varrho_s$ versus $\mu_s$  implies that the adatom mass flux is proportional to the gradient of $\exp[\mu_s/(k_BT)]$, instead of the gradient of $\mu_s/(k_B T)$ as, e.g., in~\cite{Shenoy04,BonitoNQM_09}. 

The continuum evolution law, henceforth called ``exponential PDE'', that results from the aforementioned exponential law expresses an asymmetry in the evolution of convex and concave parts of the surface. By assuming that step-step interactions are negligible, we show informally via an analytical solution that an implication of the PDE is an {\em asymmetry} in facet evolution: top and bottom facets evolve differently in a periodic surface corrugation in 1+1 dimensions. In addition, we indicate numerically how such an asymmetry manifests in the presence of elastic-dipole step-step interactions. This more complicated case lies beyond the scope of our present study.

Our approach may offer a qualitative explanation of an asymmetry in the evolution of facets of one-dimensional, periodic surface corrugations observed via kinetic Monte Carlo simulations~\cite{Zangwill92}. The authors attribute the (counter-intuitive) asymmetry in facet evolution to the relatively large amplitude of the initial height profile. Here, we view the asymmetry in facet dynamics as a direct consequence of the exponential PDE for the height profile. In this vein, we should also mention experimental observations of annealed gratings of Si with evolving facets~\cite{Tanaka97}. These observations still evade a complete understanding (see, e.g.,~\cite{IsraeliKandel00}). \color{black} Several pending questions emerge from our study. In particular, its extension to two spatial dimensions is the subject of future work.

It should be noted that in past continuum treatments of epitaxial growth, the exponential of $\mu_{\rm s}/(k_B T)$ is typically linearized under the hypothesis that $|\mu_{\rm s}|\ll k_B T$; see, e.g.,~\cite{BonitoNQM_09,KohnV,DM_Kohn06,OZ90,RV88,Shenoy02,Shenoy04}; see also the comment in~\cite{KDM}. This simplification in turn yields the standard (linear) Fick law for the mass flux in terms of the continuum-scale step chemical potential. The resulting continuum-scale evolution law does not distinguish between convex and concave parts of surface profiles.

We adopt an approach based on the following tools.  (i) The
extended-gradient (or, subgradient) formalism for the construction of
an explicit solution to the PDE for the height profile across
facets. This formalism is an extension of the PDE framework from the
previous, familiar cases of evaporation-condensation and surface diffusion under linearization of $\varrho_s$ versus $\mu_s$, in which the metric space is $L^2$ or (non-weighted) $H^{-1}$~\cite{KobayashiGiga99,Odisharia_06}, to the present, more complicated case of nonlinear gradient flow. \color{black} (ii) Numerical simulations of the PDE
by use of a regularized surface free energy, in the spirit
of~\cite{BonzelPreuss95,KohnV}. Our findings point to a few open questions about the
connection of the microscale dynamics of crystals to the corresponding
exponential PDE.

Our main results in this paper can be summarized as follows.

\begin{itemize}

\item We formulate a singular-diffusion PDE model. Away from facets, this model is consistent with the continuum limit of the Burton-Cabrera-Frank (BCF) theory for moving steps in 2+1 dimensions~\cite{BCF51,DM_Kohn06}. The PDE is also motivated by a family of KMC models of crystal surface relaxation that include both the solid-on-solid (SOS) and discrete Gaussian models~\cite{KDM,MW1}.

\item We consider the setting with a periodic surface corrugation in 1+1 dimensions, and treat facet edges as free boundaries. Accordingly, we informally develop an explicit solution for the height profile with recourse to the extended-gradient formalism in the absence of elasticity (i.e., without step-step interactions). Our construction invokes mass conservation and continuity of the continuum-scale step chemical potential across the facet. 
This procedure results in two coupled differential equations for the facet position and height, $x_f$ and $h_f$. This approach forms an extension 
of the theory underlying~\cite{GigaGiga,GigaKohn,Giga1,Odisharia_06} to the framework of exponential PDEs. 

\item 
In the context of the extended-gradient formalism outlined above, we show that the expansion of a facet is accompanied by a jump of the height profile at the facet edge; and the facet expands at finite speed.  

\item By heuristically analyzing the differential equation system for $(x_f, h_f)$ in the periodic setting without elasticity, we predict that top and bottom facets are characterized by distinctly different evolutions. In particular, the top facet starts expanding regardless of its initial size; in contrast, the bottom facet expands if its initial size exceeds a certain critical length which we compute analytically.

\item To test our analytical results, we compare them against numerical simulations by using a regularized surface free energy~\cite{BonzelPreuss95,KohnV}. Our numerics confirm our prediction that top and bottom facets behave in distinct fashion.\looseness=-1

\end{itemize}

From a physical perspective, the present, fully continuum treatment of facets, which are known to have a microscopic structure~\cite{JeongWilliams99}, leaves pending questions that need to be spelled out. The governing PDE can in principle be derived for monotone step trains; for the case of a linear-in-chemical-potential Fick law, see, e.g.,~\cite{DM_Kohn06,AlHajjShehadehKohnWeare:2011}. This type of PDE, viewed as a continuum limit of step motion, may break down in the vicinity of facets, where the distance between steps changes rapidly~\cite{IsraeliJeongKandelWeeks:2000:stepscaling,IsraeliKandel99,MFAS06}. Specifically, in the radial setting it has been shown that the continuum prediction based on the subgradient formalism may not be consistent with step flow; microscopic events of step annihilations on top of facets may significantly affect the surface slope outside the facet~\cite{MFAS06}. 

Hence, our results here are viewed as direct consequences of continuum thermodynamics and mass conservation. Our goal is to point out qualitative features of facet evolution that contrast some of the insights obtained previously by the continuum theory with a linearized law for the equilibrium adatom density versus step chemical potential. A striking feature predicted by our model is the asymmetry between top and bottom facets. 
The connection of our approach to step motion or KMC simulations in the presence of facets is left unresolved, and deserves further research in the near future. 

%%%%%%%%%%%%%%%%%%%%%%%%%%%%%%%%%%%%%%%%%%%%%%%
\subsection{Continuum framework}
\label{subsec:cont-frame}
%%%%%%%%%%%%%%%%%%%%%%%%%%%%%%%%%%%%%%%%%%%%%%%
Next, we outline the main ingredients of the continuum model in canonical form. In Section~\ref{sec:ReviewModels}, we provide details about the linkage of the continuum evolution laws to microscale models~\cite{DM_Kohn06}.

For a crystal surface evolving near a fixed crystallographic plane of symmetry, the surface free energy as a functional of height is convex and reads~\cite{GruberMullins67,Srolovitz94}
\begin{equation}
\label{eq:cont-surf-en}
E[h]=\gamma\int_\Omega \biggl(|\nabla h|+\frac{g}{3}  |\nabla h|^3\biggr)\,\dv x\qquad (\Omega\subset \mathbb{R}^2)~,
\end{equation}
where $\gamma$ is proportional to the energy cost to create a line defect (step), $h(x,t)$ is the graph of the surface, and the facet is identified with points $(x,h)$ where $\nabla h(x,t)=0$. 
% Evidently, $E[h]$ is convex. % and has a singularity at the facet with $g=0$.   
It is important that, when $g=0$, free energy~\eqref{eq:cont-surf-en} supports jumps in the height profile; see Section~\ref{sec:numerics}.   Physically, $E[h]$ expresses the joint effect of step line tension ($|\nabla h|$ term), and elastic-dipole step-step repulsive interactions
($|\nabla h|^3$ term) where $g$ is a non-negative constant equal to the relative strength of the interaction ($g\ge 0$)~\cite{GruberMullins67}; see also~\cite{Kukta02-I,Kukta02-II}. 
Formula~\eqref{eq:cont-surf-en} does not account for long-range elasticity of heteroepitaxy; see, e.g.,~\cite{Freund-book,XX}. 

Accordingly, the continuum-scale step chemical potential is defined as the variational derivative of $E[h]$, viz.,~\cite{Shenoy02}
\begin{equation}\label{eq:ms-cont}
\mu_s=\frac{\delta E}{\delta h}=-\gamma\,\diver\Biggl(\frac{\nabla h}{|\nabla h|}+g|\nabla h|\nabla h\Biggr)~,
\end{equation}
where we set the atomic volume equal to unity for algebraic convenience. Notice that~\eqref{eq:ms-cont} is ill-defined locally at the facet (where $\nabla h=0$).
By the Gibbs-Thomson relation~\cite{Rowlinson,Sethna96,DM_Kohn06}, which is connected to the theory of molecular capillarity, the corresponding local-equilibrium density of adatoms is given by
$\varrho_s=\varrho^0\,\exp[\mu_s/(k_BT)]$, where $\varrho^0$ is a constant reference density.  We note here that this relationship between the chemical potential and the density is a standard assumption, but not rigorously derived as of yet.  See \cite[Equation $(20)$]{Zangwill91} or \cite[Equations $(7)$ and $(13)$]{IhleMisbahP-L98} for further discussion. For diffusion-limited kinetics, by which  surface diffusion between steps is the rate-limiting process, by Fick's law the vector-valued adatom flux reads~\cite{DM_Kohn06}
\begin{equation}\label{eq:Fick-type}
\mathbf J=-D_s\,\nabla \varrho_s=-D_s\varrho^0\nabla e^{\mu_s/(k_B T)}~,
\end{equation}
where $D_s$ is the surface diffusion constant.

The desired evolution PDE results by combining~\eqref{eq:ms-cont} and~\eqref{eq:Fick-type} with the familiar mass conservation statement
\begin{equation}\label{eq:mass-cons}
\partial_t h+\diver \mathbf J=0~.
\end{equation}
Consequently, the height profile, $h(x,t)$, obeys the PDE
\begin{equation}\label{eq:PDE-evol}
\partial_t h=\Delta e^{-\beta\diver\left(\frac{\nabla h}{|\nabla h|}+g|\nabla h|\nabla h\right)}~,
\end{equation}
below the roughening transition; $\beta=(k_BT)^{-1}$. Here, we set the material parameter $D_s\varrho^0$ equal to unity; alternatively, this parameter, $D_s\varrho^0$, can be absorbed in the scaling of the time variable. In a similar vein, the parameter $\gamma$ is eliminated in~\eqref{eq:PDE-evol} by suitable scaling of the spatial coordinates or the Boltzmann energy, $k_B T$. \color{black} Note that a simplified version of PDE~\eqref{eq:PDE-evol} comes from linearizing the exponential of the Gibbs-Thomson relation, 
$\varrho_s\approx \varrho^0 (1+\beta\mu_s)$, under the typical assumption that the chemical potential, $\mu_s$, has magnitude sufficiently smaller than $k_B T$~\cite{Sethna96}.

%%%%%%%%%%%%%%%%%%%%%%%%%%%%%%%%%%%%%%%%%%%%%%%%%%%%%%%%
\subsection{Relevant microscopic models}
\label{subsec:micro-physics}
%%%%%%%%%%%%%%%%%%%%%%%%%%%%%%%%%%%%%%%%%%%%%%%%%%%%%%%%
The derivation of PDE~\eqref{eq:PDE-evol} is expected to hold away for facets~\cite{DM_Kohn06}. This PDE is plausibly linked to: (a) the BCF model of step flow on monotone step trains~\cite{BCF51,AlHajjShehadehKohnWeare:2011,DM_Kohn06}; and (b) a family of atomistic models~\cite{MW1}. Here, we outline elements of these microscale theories. In Section~\ref{sec:ReviewModels}, we provide a more detailed review of their linkages to~\eqref{eq:PDE-evol}.

First, consider the mesoscale picture of step flow. The BCF model accounts for diffusion of adatoms and attachment/detachment of atoms at steps~\cite{BCF51}. Key ingredients of the respective formalism are: (i) a step velocity law by mass conservation; (ii) a diffusion equation for adatoms on each nanoscale domain, terrace, between steps; and (iii) a Robin boundary condition for the adatom density at the step edge. Hence, the step is viewed as a free boundary  for a Stefan-type problem; the step position is determined via diffusion and each terrace is a level set for the height. In the kinetic regime of diffusion-limited kinetics, the Robin boundary condition reduces to a Dirichlet condition~\cite{BCF51}. In the continuum limit, the step height, which is equal to the vertical lattice spacing, approaches zero while the surface slope is kept fixed.

Alternatively, in the respective atomistic picture based on the SOS model, the core mechanism is the hopping of atoms on the crystal lattice~\cite{Weeks79,MW1}. The formalism relies on a Markovian process representing
the motion of each atom from one lattice site to a neighboring site. In this model, the transitions between atomistic configurations are determined by Arrhenius rates which in turn are related to the number of bonds that each atom would be required to break in order to move. In~\cite{MW1}, a macroscopic limit of these dynamics, as the lattice spacing vanishes, is proposed via the form of the surface tension as the $p$-Laplacian for the potential $V(x)=|x|^p$, $p>1$. 
%and should include a coarser definition of the surface tension in the initial dynamics %of the smooth crystal.
PDE~\eqref{eq:PDE-evol} is an extension of that macroscopic limit in~\cite{MW1} to $p=1$. Notably, the resulting PDE is sensitive to the way by which the initial height profile is scaled~\cite{MW1}.

%%%%%%%%%%%%%%%%%%%%%%%%%%%%%%%%%%%%%%%%%%%%%%%%%%%%%%%%%%%%%%
\subsection{Our mathematical approach and core result}
\label{subsec:approach}
%%%%%%%%%%%%%%%%%%%%%%%%%%%%%%%%%%%%%%%%%%%%%%%%%%%%%%%%%%%%%%
Our mathematical approach makes use of a version of the subgradient formalism~\cite{KobayashiGiga99}, adapted to the exponential, fourth-order PDE~\eqref{eq:PDE-evol}. In physical terms, intuitively, this formalism may be viewed as tantamount to a limiting procedure by which the facet is artificially smoothed out and then is allowed to approach a flat plateau. This procedure can be viewed as the outcome of the regularization of the 
surface free energy, $E[h]$; see, e.g.,~\cite{BonzelPreuss95,KohnV}. It should be noted that a different approach of regularization found in the literature relies on the truncation of Fourier series expansions for the height 
profile, which yields nonlinear differential equations for the requisite coefficients~\cite{Shenoy02,Shenoy04,Chan04}.

Our construction of a solution treats the facet edge as a free boundary, in the spirit of~\cite{Spohn93}. In the continuum thermodynamics framework, the boundary conditions at the facet edge result from the 
%abstraction by the 
extended-gradient formalism as follows. Replace PDE~\eqref{eq:PDE-evol} by the statement that $\partial_t h$ {\em picks} the subgradient of $E[h]$ with the minimal norm in the appropriate metric; see, e.g.,~\cite{Odisharia_06,Kashima} for works on the $H^{-1}$ gradient flow. 
In the present case, in $1+1$ dimensions PDE~\eqref{eq:PDE-evol} is replaced by the statement that \eqref{eq:PDE-evol} with $g=0$ can be realized as the nonlinear $H^{-1}$ flow given by
\begin{equation}\notag
 \begin{aligned}
  \partial_t h & = - \partial_x \left[e^{- \partial_x\left(\frac{ \partial_x h }{ |\partial_x h|}\right)  } \partial_{xx} \left(  \frac{ \partial_x h }{ |\partial_x h|}   \right)\right] \\
  & =  \partial_x\left[ e^{- \partial_x\left(\frac{ \partial_x h }{ |\partial_x h|} \right) } \partial_x \left( \frac{\delta E}{ \delta h}  \right)\right]~.
  \end{aligned}
\end{equation}
  This evolution can be viewed as a nonlinear $H^{-1}$ gradient flow with (exponential) mobility equal to $e^{- \partial_x\left(\frac{ \partial_x h }{ |\partial_x h|} \right) }$.  We write
  \begin{equation}\notag
\partial_t h= \partial_{xxx} v\quad \mbox{where}
\quad \partial_x v=e^{-\partial_x w}~,
\end{equation}
where  $w = h_x/|h_x|$, $-\partial_x w \in \partial_{L^2} E$ is an element of the $L^2$-subdifferential of $E[h]$, and the function $v(x)$ is determined in the sense described in Section~\ref{sec:ODE-facets}. The functions $v$ and $\partial_x v$ \color{black} are continuous; in addition, these functions are subject to the symmetry of the surface profile. Thus, $-\partial_x w=\mu$, the continuum-scale step chemical potential, and $w$ are continuous across the facet. Furthermore, the mass conservation statement  $\partial_t h+\partial_x J=0$ where $J=-\partial_{xx}v$ is the $x$-component of the (vector-valued) adatom flux $\mathbf J$, entails a jump condition for the continuum-scale adatom flux and height across the facet edge~\cite{GigaGiga}. It should be borne in mind that the facet height, $h_f$, is constant in $x$; thus, the above conditions can be applied by successive integrations with respect to $x$ of the conservation law for the height, where $\partial_t h$ in the facet region is the vertical facet speed, $\dot h_f$.

For $g=0$, i.e., if step-step interactions are neglected, this procedure entails a  discontinuous height and mass flux at the facet boundary, in agreement with rigorous studies in \cite{GigaGiga} on the total variation flow model 
\begin{equation}
\label{eq:gigatvpde}
\partial_t h = -\partial_x^3\left(\frac{\partial_x h}{| \partial_x h|} \right) =  \partial_x^2 \biggl( \frac{ \delta E}{ \delta h} \biggr)~,
\end{equation}
which has the structure of a (non-weighted) $H^{-1}$ gradient flow. The reader is referred to~\cite{FLL,XX} for related works in the presence of elasticity.

A noteworthy result here is the derivation of a system of two differential equations for the facet position, $x_f$, and facet height, $h_f$, via the exponential PDE. By properties of this system, we infer that facets in convex and concave parts of the surface {\em behave differently}. In particular, by our theory, if the initial height profile is sinusoidal, the surface peaks immediately break into expanding facets; in contrast, no facets form at the valleys of the initial profile (see Section~\ref{sec:numerics}). It should be mentioned that experimental observations in epitaxial relaxation do not seem to report the formation of asymmetric facets of one-dimensional corrugations, although lack of symmetry in facet dynamics is observed in two dimensions in a certain temperature range~\cite{Tanaka97}. \color{black}

%%%%%%%%%%%%%%%%%%%%%%%%%%%%%%%%%%%%%%%%%%%%%%%%%%%%%%%%%%%%%
\subsection{Limitations}
\label{subsec:limitations}
%%%%%%%%%%%%%%%%%%%%%%%%%%%%%%%%%%%%%%%%%%%%%%%%%%%%%%%%%%%%%
Our work points to several open questions. First, the precise nature of the gradient flow for PDE~\eqref{eq:PDE-evol} is not adequately understood. \color{black} 
As noted earlier, the comparison of our continuum predictions to results 
from the step flow near facets is an interesting problem left for future research. 
A requisite issue in this context is the sign of the interactions between 
colliding steps on facets~\cite{IsraeliKandel00}. In a similar vein, we do not 
pursue comparisons of the continuum predictions against KMC simulations, which 
would connect the PDE solution to atomistic dynamics; see~\cite{MW1}. Our construction 
of an explicit solution to the exponential PDE focuses on one spatial coordinate with 
diffusion-limited kinetics and $g=0$. In 2+1 dimensions or settings with elasticity 
or other kinetics (say, attachment-detachment limited kinetics), 
the subgradient formalism becomes more intricate. The facet evolution in such cases 
needs to be further studied.

%%%%%%%%%%%%%%%%%%%%%%%%%%%%%%%%%%%%%%%%%%%%
\subsection{Paper outline}
\label{subsec:outline}
%%%%%%%%%%%%%%%%%%%%%%%%%%%%%%%%%%%%%%%%%%%%%
The remainder of this paper is organized as follows. In Section~\ref{sec:ReviewModels}, we review linkages of PDE~\eqref{eq:PDE-evol} to existing microscopic models.  Section~\ref{sec:ODE-facets} focuses on the construction of the ODEs governing facet dynamics.  In Section~\ref{sec:numerics}, we numerically solve both the ODE system and an appropriately regularized version of the PDE; and compare the outcomes. Finally, in Section~\ref{sec:Discussion}, we summarize the results obtained and outline some topics for future work.

%%%%%%%%%%%%%%%%%%%%%%%%%%%%%%%%%%%%%%%%%%%%%%%%%%%%%%%%%%%%
%%%%%%%%%%%%%%%%%%%%%%%%%%%%%%%%%%%%%%%%%%%%%%%%%%%%%%%%%%%%
\section{Mesoscale and atomistic descriptions: A review} 
\label{sec:ReviewModels}
%%%%%%%%%%%%%%%%%%%%%%%%%%%%%%%%%%%%%%%%%%%%%%%%%%%%%%%%%%%%
%%%%%%%%%%%%%%%%%%%%%%%%%%%%%%%%%%%%%%%%%%%%%%%%%%%%%%%%%%%%

In this section, we describe ingredients of the mesoscale and atomistic models that motivate the study of~\eqref{eq:PDE-evol} as a hydrodynamic-type limit. In particular, we review basics of the BCF model~\cite{BCF51} and a heuristic derivation of its continuum limit assuming that this limit exists (Section~\ref{subsec:BCF-limit}). We also outline the relevance to the exponential PDE of a kinetic Monte Carlo model of crystal surface relaxation~\cite{MW1} (Section~\ref{subsec:broken-bond-limit}). The emergence of the BCF description of step flow from atomistic dynamics is not addressed here; see, e.g.,~\cite{LLM}.

%%%%%%%%%%%%%%%%%%%%%%%%%%%%%%%%%%%%%%%%%%%%
\subsection{BCF model and its continuum limit}
\label{subsec:BCF-limit}
%%%%%%%%%%%%%%%%%%%%%%%%%%%%%%%%%%%%%%%%%%%%
By the BCF model~\cite{BCF51}, the crystal surface consists of atomic steps separated by nanoscale terraces.  In this subsection, we review the basic elements of step flow, needed for our purposes, by mainly following the formalism of~\cite{DM_Kohn06}.

\begin{figure}
\begin{center}
\includegraphics[scale=0.7]{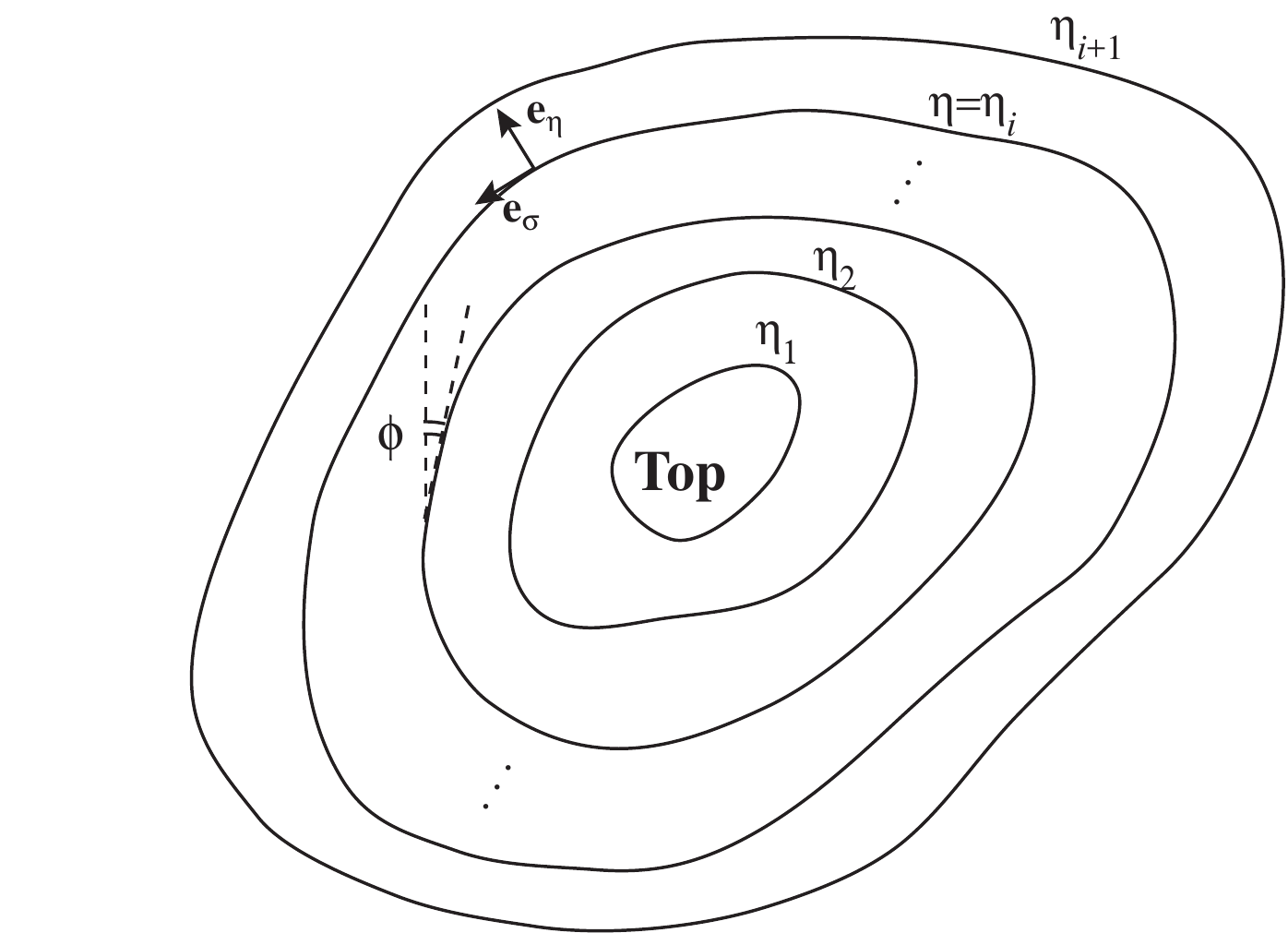}
\caption{Geometry of monotone step train in 2+1 dimensions (top view). In curvilinear coordinates $(\eta, \sigma)$, the depicted contours are projections of descending steps onto a fixed reference plane; $\eta=\eta_i$ at the $i$-th edge while $\sigma$ varies along a step edge. 
The step orientation, relative to a fixed axis, is indicated by the (local) angle $\phi$.}
\label{f:step-geometry}
\end{center}
\end{figure}

Figure~\ref{f:step-geometry} depicts the top view of descending, non-intersecting steps of atomic height $a$ in two dimensions. The steps surround a top terrace. The projections of the step edges onto a fixed, high-symmetry plane of reference are modeled by a family of smooth curves, numbered by $i$ ($i=1,\,2,\,\ldots, N$) relative to the top terrace where $N\gg 1$. These curves are described by the position vector 
    $\mathbf r(\eta,\sigma; t)$. The variable $\eta$ corresponds to the polar coordinate for the distance from the origin of the radial case in which the steps are concentric circles; in general, $\eta=\eta_i$ for the $i$-th edge and $\eta_i<\eta<\eta_{i+1}$
    for the $i$-th terrace. The variable $\sigma$ corresponds to the angle in polar coordinates and increases counterlockwise ($0\le\sigma < 2\pi$). \color{black} Thus, $\eta$ identifies each step and $\sigma$ specifies the position along a step edge. The unit vectors normal and parallel to step edges in the direction of increasing $\eta$ and $\sigma$ are denoted $\mathbf e_\eta$ and 
$\mathbf e_\sigma$, where we take $\mathbf e_\eta \cdot \mathbf e_\sigma=0$. The corresponding metric coefficients are 
\begin{equation}\label{eq:metric-coeffs}
\xi_\eta=|\partial_\eta \mathbf r|~,\quad \xi_\sigma=|\partial_\sigma \mathbf r|~.
\end{equation}
The $i$-th step is the  set $\{(\eta,\sigma)\,:\,\eta=\eta_i={\rm const.}\}$, and the $i$-th terrace is the region $\{(\eta,\sigma)\,:\,\eta_i <\eta <\eta_{i+1}\}$.  Hence, the surface height, $h$, is a function of $\eta$ only, and obeys
\begin{equation}\notag
h|_{\eta=\eta_{i+1}}-h|_{\eta=\eta_i}=-a~.
\end{equation}
In the continuum limit, we let $a\downarrow 0$ while we keep the step density fixed. The Taylor expansion of the left-hand side of the last equation entails that $a/(\xi_\eta\delta\eta_i)$ approaches the (fixed)  positive slope
$-(\partial_\perp h)|_{\eta=\eta_i}=|\nabla h|$ as $a\downarrow 0$ and $\delta\eta_i:=\eta_{i+1}-\eta_i\downarrow 0$; note that $\partial_\perp:=\xi_\eta^{-1}\partial_\eta$.  

%%%%%%%%%%%%%%%%%%%%%%%%%%%%%%%%%%%%%%%%%%%%%%%%%%%%%%%%%%%%
\subsubsection{Laws of step flow and continuum limit}
\label{sssec:step-cont}
%%%%%%%%%%%%%%%%%%%%%%%%%%%%%%%%%%%%%%%%%%%%%%%%%%%%%%%%%%%%
First, the normal velocity of the $i$-th step is given by
\begin{equation}\label{eq:n-step-vel}
v_{i,\perp}= a^{-1}(J_{i-1,\perp}-J_{i,\perp})\qquad \mbox{at}\ (\eta_i,\sigma)~.
\end{equation}
Here, $J_{i,\perp}=\mathbf e_\eta\cdot \mathbf J_i$ where $\mathbf J_i=-D_s \nabla \varrho_i$ is the 
vector-valued adatom flux on the $i$-th terrace and $\varrho_i$ is the respective adatom concentration. In the quasi-steady approximation, ${\rm div}\mathbf J_i\approx 0$ on the $i$-th terrace. 

In the continuum limit, as $a\downarrow 0$, \eqref{eq:n-step-vel} reduces to a mass conservation statement.
Indeed, the $v_{i,\perp}$ approaches $\partial_t h/|\nabla h|$ at $(\eta_i,\sigma)$. Furthermore, by Taylor expanding we have
$J_{i,\perp}(\eta_i,\sigma)\approx J_{i,\perp}(\eta_{i+1},\sigma)-(\delta\eta_i)\partial_\eta J_{i,\perp}(\eta_i,\sigma)=J_{i,\perp}(\eta_{i+1},\sigma)+(\xi_\eta\delta\eta_i)\xi_\sigma^{-1}\partial_\sigma J_{i,\parallel}$
where $J_{i,\parallel}=\mathbf e_\sigma\cdot \mathbf J_i(\eta_i,\sigma)$ and use was made of ${\rm div}\mathbf J_i\approx 0$.
Thus, the right-hand side of~\eqref{eq:n-step-vel} approximately reads 
$-a^{-1}(\delta\eta_i)\partial_\eta J_\perp-a^{-1}(\xi_\eta\delta\eta_i)\xi_\sigma^{-1}\partial_\sigma J_{\parallel}$ which is identified with $-|\nabla h|^{-1}{\rm div}\mathbf J$; $\mathbf J(\mathbf x,t)$ is the continuum-scale adatom flux, with $J_\perp=\mathbf e_\eta\cdot \mathbf J$ and $J_\parallel=\mathbf e_\sigma\cdot \mathbf J$. Therefore, we obtain
\begin{equation}
	\partial_t h=-{\rm div}\mathbf J~.
\end{equation}

Next, we consider the attachment/detachment of atoms at steps. By the quasi-steady approximation, we set $D_s\Delta\varrho_i=\partial_t\varrho_i\approx 0$ on the $i$-th terrace. The boundary conditions for this diffusion equation are of the Robin type, viz.,
\begin{equation}\label{eq:bc-at}
	-J_{i,\perp}=k (\varrho_i^+ -\varrho^{\rm eq}_i)\quad \mbox{at}\ (\eta_i,\sigma)~,\qquad J_{i,\perp}=k(\varrho_{i}^--\varrho_{i+1}^{\rm eq})\quad \mbox{at}\ (\eta_{i+1},\sigma')~,
\end{equation}
where $\varrho_i^{\pm}$ is the restriction of $\varrho_i$ at a step edge as $\eta$ approaches: $\eta_i$ ($+$), or $\eta_{i+1}$ ($-$) on the $i$-th terrace. The quantity $\varrho_i^{\rm eq}$ is the equilibrium adatom density at the $i$th step and is given by the Gibbs-Thomson relation (discussed below). Equations~\eqref{eq:bc-at} are combined for $\sigma\neq \sigma'$ to yield
\begin{equation}\notag
J_{i,\perp}(\eta_i,\sigma)+J_{i,\perp}(\eta_{i+1},\sigma')=	k[\varrho_i(\eta_{i+1},\sigma')-\varrho_i(\eta_i,\sigma)]-k[\varrho_{i+1}^{\rm eq}(\sigma')-\varrho_i^{\rm eq}(\sigma)]~.
\end{equation}
We now show that, in the limit where $\delta\sigma:=\sigma'-\sigma\to 0$ and $\delta\eta_i\downarrow 0$, the last equation entails a Fick-type law for $\mathbf J$ in terms of the continuum-scale equilibrium density, $\varrho^{\rm eq}$. Notice that $\delta\eta_i$ is $\mathcal O(a)$, because the slope is kept fixed, whereas $\delta\sigma$ is allowed to approach zero independently of $a$. By assuming that
\begin{equation}\notag
\frac{D_s}{ka}=\mathcal O(1)~,	
\end{equation}
consider the Taylor expansions 
\begin{equation}
\begin{aligned}
\varrho_i(\eta_{i+1},\sigma')-\varrho_i(\eta_i,\sigma)&= (\delta\eta_i)\,\partial_\eta\varrho_i+(\delta\sigma)\partial_\sigma\varrho_i+\mathcal O((\delta\eta_i)^2 + (\delta\sigma)^2)\\
&\approx -(\xi_\eta\delta\eta_i)D_s^{-1}J_{i,\perp}-(\xi_\sigma\delta\sigma)D_s^{-1}J_{i,\parallel}
\end{aligned}
\end{equation}
and 
\begin{equation}\notag
	\varrho_{i+1}^{\rm eq}(\sigma')-\varrho_i^{\rm eq}(\sigma)\approx (\xi_\eta\delta\eta_i)\partial_\perp\varrho^{\rm eq}+(\xi_\sigma\delta\sigma)\partial_\parallel\varrho^{\rm eq}~,\quad \partial_\parallel:=\xi_\sigma^{-1}\partial_\sigma~.
\end{equation}
Accordingly, we obtain the expression
\begin{multline}\notag
(\xi_\eta\delta\eta)\left\{\left(\frac{2D_s}{ka}\,|\nabla h|+1\right)J_\perp
+D_s\,\partial_\perp\varrho^{\rm eq}\right\}+\,(\xi_\sigma\delta\sigma)\left\{J_\parallel+D_s\,\partial_\parallel\varrho^{\rm eq}\right\}\\
=\mathcal O(a^2,a(\xi_\sigma\delta\sigma)) 
\end{multline}
at the point $(\eta_i,\sigma)$, provided $|\nabla h|=\mathcal O(1)$ ($|\nabla h|\neq 0$). Hence, by setting each term equal to zero in the first line, we extract the formulas
\begin{subequations}\label{eq:Fick-law}
\begin{equation}\label{eq:Fick-law-gen}
J_\perp=-\frac{D_s}{{\displaystyle 1+\frac{2D_s}{ka}|\nabla h|}}\,\partial_\perp\varrho^{\rm eq}~,\quad J_\parallel=-D_s\,\partial_\parallel \varrho^{\rm eq}~,	
\end{equation}
in the local coordinate system. In particular, for diffusion-limited kinetics, when the diffusion of adatoms on terraces is the slowest process, the length 
$D_s/k$ is much smaller than the terrace size, $[D_s/(ka)]|\nabla h|\ll 1$; thus, we find 
\begin{equation}\label{eq:Fick-law-DL}
\mathbf J=-D_s\,\nabla\varrho^{\rm eq}\quad \mbox{if}\quad \frac{D_s}{ka}|\nabla h|\ll 1~;
\end{equation}
\end{subequations}
cf.~\eqref{eq:Fick-type} if $\varrho_s$ is identified with $\varrho^{\rm eq}$.   

Equations~\eqref{eq:mass-cons} and~\eqref{eq:Fick-law} need to be complemented with a formula for $\varrho^{\rm eq}$ involving the continuum-scale step chemical potential, $\mu_s$. At the level of step flow, the Gibbs-Thomson relation dictates that
\begin{equation}\label{eq:Gibbs-Thomson-r}
\varrho_i^{\rm eq}=	\varrho^0 e^{\frac{\mu_i}{k_BT}}~,
\end{equation}
where $\varrho^0$ is a reference density for an atomically flat terrace. The step chemical potential, $\mu_i(\sigma)$, of the $i$-th step is defined as the  change of the step energy by addition or removal of an atom to or from the step edge at $\eta=\eta_i$. Following~\cite{DM_Kohn06}, consider a short step length, $\dv s =\xi_\sigma \dv \sigma$, of the $i$-th edge that has energy 
$\mathcal U_i \dv s$ at $(\eta_i,\sigma)$; $\mathcal U_i$ is the step energy per unit length. The exchange of atoms with the step edge results in the motion of the step along its local normal by distance 
$\dv r=\xi_\eta \dv\eta$ where $\dv\eta$ is the respective shift of $\eta_i$. Hence, the step energy $\mathcal U_i\dv s$ changes by
$\dv_\eta(\mathcal U_i\,\dv \sigma)$, where the shift operator $\dv_\eta$ is defined by $\dv_\eta Q := Q|_{\eta+\dv\eta} - Q|_{\eta}$. Accordingly, we write
\begin{equation}\label{eq:chem-pot-i}
	\mu_i=\frac{1}{a}\frac{\dv_\eta(\mathcal U_i\dv s)}{\dv r\dv s}=\frac{1}{a}\{\xi_\eta^{-1}\partial_{\eta_i}\mathcal U_i+\mathcal U_i\,(\xi_\eta\xi_\sigma)^{-1}\partial_\eta\xi_\sigma\}\quad \mbox{at}\ \eta=\eta_i~.
\end{equation}
By using the elementary formula $\xi_\eta^{-1}\partial_\eta\xi_\sigma=\kappa\xi_\sigma$ where $\kappa$ is the curvature of the curve $\mathbf r(\eta,\sigma;t)$ with $\eta={\rm const.}$, we obtain
\begin{equation}\label{eq:mu-i}
\mu_i=\frac{1}{a}\left(\kappa_i \mathcal U_i+\xi_{\eta_i}^{-1}\partial_{\eta_i}\mathcal U_i\right)~.
	\end{equation}
	The quantity $\mathcal U_i$ incorporates the step line tension, $\tilde\gamma_i$, which is the energy cost per unit length to create a step and may in principle depend on the step orientation $\phi$ (see Figure~\ref{f:step-geometry}), \color{black} as well as the step-step interaction contribution, $\mathcal U_i^{\rm int}$.
	In a simple scenario for homoepitaxy, $\tilde \gamma_i=\gamma a$ is a global, material-dependent constant; and step interactions are modeled as nearest-neighbor repulsions~\cite{MarchenkoP80,Srolovitz94}, viz.,
	\begin{equation}
	\mathcal U_i=a\gamma+\mathcal U_i^{\rm int}~,\quad \mathcal U_i^{\rm int}=\tilde g\,(V_{i,i+1}+V_{i,i-1})~,
	\end{equation}
	where $\tilde g$ is the interaction strength (energy/length), and 
	$V_{i,i\pm 1}$ amounts to the interaction between the $i$-th and 
	$(i \pm 1)$-th steps and depends on $\eta_i$ and $\eta_{i\pm 1}$. For elastic-dipole or entropic interactions, the $V_{i,j}$ ($j=i\pm 1$) is taken to be~\cite{DM_Kohn06}
	\begin{equation}\notag
		V_{i,i+1}=\frac{1}{3}m_i^2\Phi(r_i,r_{i+1})~,\quad V_{i,i-1}=\frac{1}{3}m_{i-1}^2\Phi(r_i,r_{i-1})~,
	\end{equation}
	where
	\begin{equation}\notag
		m_i:=\frac{a}{r_{i+1}-r_i}~;\quad r_i=r|_{\eta=\eta_i}~,\quad r=\int_{\eta_0}^\eta \xi_{\eta'}\,\dv\eta'~,
	\end{equation}
	and $\Phi(\zeta,\chi)$ is a geometrical factor described in some detail in~\cite{DM_Kohn06}.
	
	In the continuum limit, the curvature $\kappa_i$ of the step approaches 
	$-{\rm div}[\nabla h/|\nabla h|]$ at the point $(\eta_i,\sigma)$. By~\eqref{eq:mu-i},
	the step chemical potential, $\mu_i$, approaches form~\eqref{eq:ms-cont} under mild assumptions for $\Phi$. The interested reader is referred to~\cite{DM_Kohn06}.

	  Note that in the attachment-detachment-limited regime, where 
$$
\frac{D_s}{ka}|\nabla h|\gg 1~,
$$
the attachment/detachment of atoms at steps is the slowest process. In this case, the resulting PDE in 1+1 dimensions acquires a slope-dependent, extra mobility, viz.,
\begin{equation}
\partial_t h=\partial_x \left[\frac{1}{|\partial_x h|}\partial_x e^{-\beta\partial_x\left(\frac{\partial_x h}{|\partial_x h|}+g|\partial_x h|\partial_x h\right)}\right]~.	
\end{equation}
The study of this PDE lies beyond our present scope.

%%%%%%%%%%%%%%%%%%%%%%%%%%%%%%%%%%%%%%%%%%%%%%%%%%%%%%
\subsection{Broken-bond model and hydrodynamic limit}
\label{subsec:broken-bond-limit}
%%%%%%%%%%%%%%%%%%%%%%%%%%%%%%%%%%%%%%%%%%%%%%%%%%%%%%
In this subsection, we review aspects of the emergence of continuum laws from atomistic principles in~\cite{MW1}.
Motivated by an adatom model proposed in \cite{KDM} and studies of hydrodynamic limits undertaken in~\cite{Funaki,funakispohn,Nishikawa}, the authors in \cite{MW1} derive exponential PDEs of form similar to \eqref{eq:PDE-evol}.  
This atomistic formulation views the crystal surface as a function  $h_N(\alpha, t)$  for time $t\ge 0$ and position $\alpha\in \mathbb{T}^d_N = \left( \mathbb{Z}/ N\mathbb{Z}\right)^d$ on the periodic lattice,
 with  values of $h_N$ in the set of integers, $\mathbb{Z}$; $d$ is the spatial dimension.  
The rates are related to an interaction potential, $V: \mathbb{Z} \rightarrow \mathbb{R}$, taken to be the non-negative, strictly convex, symmetric function, $V(z) = |z|^p$, of the discrete slope, $z$. The choice for $V$ made in \cite{MW1}, and the most common choice in the literature on the physics of crystal surfaces, is $V(z) = |z|$ (if $p=1$), which amounts to bond breaking by the SOS model~\cite{Weeks79}.

From such an interaction potential, $V=|z|^p$, in \cite{MW1} a family of Arrhenius rates are proposed based upon a generalized coordination number.  One can think of the generalized coordination number as the (symmetrized) energy cost associated with removing a single atom from site $\alpha$ on the crystal surface, where the energy is determined by summing over the interaction potential evaluated on local fluxes.  
%PDEs with exponential mobility are derived using a rough scaling limit of the form $\bar{f}_N (t,x) = N^{-q} f( N^{q+2} t, \alpha)$ with $Nx \in [\alpha-1/2, \alpha + 1/2)$ with $q = p/(p-1)$, where the crystal surface energy is determined by an interaction potential.

In \cite{MW1}, two scaling limits are studied.  First, for $p \geq 1$, the evolution
of the height of a smooth crystal surface is found to be
\begin{equation}
\label{eqn:krug1}
\partial_t h = -\frac{1}{2d}  \Delta ( \text{div} ( \nabla \sigma_D ( \nabla h))~,
\end{equation}
where $ \nabla \sigma_D : \RR^d \to \RR^d$ is the gradient of the surface tension, $\sigma_D$, which is a convex function determined 
by a free-energy computation and depends on the choice of the interaction potential, $V$.  The definition of this $\sigma_D$ arises from essentially applying the local Gibbs measure (local equilibrium) for finding non-equilibrium dynamics in the microscopic model of crystal surface evolution~\cite{MW1}.  In particular, $\sigma_D$ stems from using a discrete chemical potential, which matches well the macroscopic dynamics; see \cite{MW1}.  PDE~\eqref{eqn:krug1} arises from a smooth diffusion scaling limit of the form $\bar{h}_N (x,t) = N^{-1} h(  \alpha, N^{4} t)$ with $Nx \in [\alpha-1/2, \alpha + 1/2)$.  

Similarly, in \cite{MW1}, a second PDE for a rough crystal evolution is proposed for $p > 1$ with fixed temperature $\beta^{-1}$ ($\beta>0$), viz.,
\begin{equation}
\label{eqn:krug2}
  \partial_t h = \frac{1}{2d} \Delta\left( e^{- \text{div}\left( \nabla  \sigma_C(\nabla \bar h)\right)} \right) \ ; \quad
\sigma_C (z) = \lim_{\kappa \rightarrow \infty} \kappa^{-p} \sigma_D (\kappa z)~.
\end{equation}
The form that the surface tension then takes is the $p$-Laplacian for $ \sigma_C (z) = \beta |z|^p$, resulting in the evolution
\begin{equation}
\label{eqn:krug2p}
    \partial_t h = \frac{1}{2d} \Delta\left( e^{- \beta \text{div}\left( (|\nabla h|^{p-2} \nabla h) \ \right)} \right)~.
\end{equation}
 This PDE arises from a rough scaling limit of the form $\bar{h}_N (x,t) = N^{-q} h( \alpha,  N^{q+2} t)$ with $Nx \in [\alpha-1/2, \alpha + 1/2)$ and $q = p/(p-1)$.  

However, the rough scaling when $p=1$ can be adapted by formally following the derivation in \cite[Section $6.2$]{MW1}, if one systematically lowers the temperature, $\beta^{-1}$, as one increases the system size ($\beta = \beta (N)$ such that $\beta (N) \to \infty$ as $N \to \infty$).  Then, the methods of \cite{MW1} can be invoked to derive \eqref{eq:PDE-evol} with Boltzmann constant $\tilde \beta$, and, for instance, $q = 1$ and $\beta (N) = \tilde \beta N$.

%%%%%%%%%%%%%%%%%%%%%%%%%%%%%%%%%%%%%%%%%%%%%%%%%%%%%%
%%%%%%%%%%%%%%%%%%%%%%%%%%%%%%%%%%%%%%%%%%%%%%%%%%%%%%
\section{ODE system for facet motion via exponential PDE}
\label{sec:ODE-facets}
%%%%%%%%%%%%%%%%%%%%%%%%%%%%%%%%%%%%%%%%%%%%%%%%%%%%%%
%%%%%%%%%%%%%%%%%%%%%%%%%%%%%%%%%%%%%%%%%%%%%%%%%%%%%%
In this section, we formulate an ODE system for facets in a periodic surface corrugation with $g=0$ (no elasticity) \color{black} in 1+1 dimensions. This amounts to the construction of a solution for the height profile. We first discuss a general framework for the gradient flow. \color{black} Accordingly, we analytically indicate the different behaviors of top and bottom facets by use of the ODEs. For algebraic convenience, we use PDE~\eqref{eq:PDE-evol} by setting $\beta$ equal to unity, absorbing $\beta$ into the spatial coordinates. In Section~\ref{sec:numerics}, our construction of a solution is compared to numerics for the PDE by regularization of the surface free energy, $E[h]$.\color{black}

In the case with the (non-weighted) $H^{-1}$ total variation flow (e.g. \cite{GigaGiga,GigaKohn,Giga1}), the PDE takes the form
%In \cite{GigaGiga}, the authors have studied the $H^{-1}$ total variation flow of the form
\begin{equation}
\label{gigapde}
\partial_t h = -\partial_{xx} \partial_x \left( \frac{\partial_x h}{ | \partial_x h |} \right), \ \ h(x,0) = h_0 (x)~,
\end{equation}
where $h_0 (x)$ is assumed to be symmetric and \color{black} have an extremum at $x=0$.  A weak solution to \eqref{gigapde} is derived in \cite{GigaGiga} as a facet solution (symmetric about the maximum or minimum of $h_0$ \color{black} at $x=0$) near the critical point $x=0$ of $h_0$. This weak solution has the form 
\begin{equation}\notag
h(x,t) = 
\begin{cases}
h_f (t) &  \text{for} \ x < x_f (t)~, \\
h_0 (x)  &  \text{for} \ x > x_f (t)~,
\end{cases}
\end{equation}
where $x=x_f(t)$ is the facet position and $h_f(t)$ is the facet height.
The dynamics for $(x_f (t), h_f(t))$ obey the ODE system
\begin{equation}
\left\{  \begin{array}{l}
\dot h_f (t) = - \frac{3}{ x_f^3 (t) }~,   \\
 \dot x_f (t)  ( h_0 (x_f) - h_f (t) ) = -3 x_f^{-2}(t)~.
 \end{array} \right. \label{gigasys}
\end{equation}
This is the symmetric formulation for the $H^{-1}$ total variation flow.  For the case of the $L^2$ total variation flow, in which the PDE for $h$ is of second order, see,  e.g., \cite{KobayashiGiga99}.

We now turn our attention to exponential PDE~\eqref{eq:PDE-evol} with $g=0$. For this case, we lack a mathematically rigorous theory. \color{black} Following the works of \cite{GigaGiga,GigaKohn,Giga1}, we recognize that \eqref{eq:PDE-evol} with $g=0$ can be realized as the (nonlinear) $H^{-1}$ flow given by
\begin{equation}\notag
 \begin{aligned}
  \partial_t h & = - \partial_x \left[e^{- \partial_x\left(\frac{ \partial_x h }{ |\partial_x h|}\right)  } \partial_{xx} \left(  \frac{ \partial_x h }{ |\partial_x h|}   \right)\right] \\
  & =  \partial_x\left[ e^{- \partial_x\left(\frac{ \partial_x h }{ |\partial_x h|} \right) } \partial_x \left( \frac{\delta E}{ \delta h}  \right)\right]~.
  \end{aligned}
\end{equation}
  This evolution is viewed as a nonlinear (weighted) $H^{-1}$ flow with mobility $e^{- \partial_x\left(\frac{ \partial_x h }{ |\partial_x h|} \right) }$, as we explain below (see Section~\ref{subsec:grad-glow}). \color{black} {\it This model of evolution implies an asymmetry between convex and concave parts of the crystal surface.}  
  
We now discuss aspects of the subdifferential (see \cite{GigaGiga,Kashima}) associated with $E[h]$ in our setting, by assuming a symmetric height profile. Recall that the facet is the set of points $(x,h)$ with $\partial_x h(x,t)=0$. Let us introduce the functional $F: D \rightarrow \mathbb{R}$ by
\begin{equation*}
	F[v]:=\int_{-r_0}^{r_0} (\partial_{xx}v)^2\,\dv x~,
\end{equation*}
with domain $D(F)=\{v\in H^2[-r_0,r_0]:\, v\ \mbox{is\ odd}\ \mbox{and}\  \partial_x v (\pm r_0) = 1 \}$ where $-r_0\le x\le r_0$. We will set $\partial_x v=e^\mu$; thus, the condition $\partial_x v(\pm r_0)=1$ accounts for the effective values $\pm 1$ that $\frac{\partial_x h }{ |\partial_x h|}$ has outside the facet (if $\partial_x h\neq 0$). We claim that the desired element $-\partial_x w\in \partial_{L^2}E$ in evolution is such that 
$\partial_x v^* = e^{-\partial_x w}$
 where $v^*\in D(F)$ is a minimizer of $F$. Note that outside the facet $w(u) = u/ |u|$ where $u$ is identified with the slope profile, $\partial_x h$.  In the above, $H^2$ denotes the metric (Sobolev) space of all locally summable functions whose derivatives of order less or equal to two exist in the weak sense and are square integrable in the domain. \color{black}   
 
 %%%%%%%%%%%%%%%%%%%%%%%%%%%%%%%%%%%%%%%%%%%%%%%%%%%%%%%%%%%%%%%%%%%%%%%%%%%%%%%%%%%%%%%%%%%%%%%%
 \subsection{On the machinery of the gradient flow}
 \label{subsec:grad-glow}
 %%%%%%%%%%%%%%%%%%%%%%%%%%%%%%%%%%%%%%%%%%%%%%%%%%%%%%%%%%%%%%%%%%%%%%%%%%%%%%%%%%%%%%%%%%%%%%%% 
Next, we discuss the meaning of the gradient flow for PDE~\eqref{eq:PDE-evol} with $g=0$ by recourse to Ambrosio, Gigli and Satar\'e \cite{AGS}. Our goal is to further illuminate the subdifferential structure underlying this PDE. \color{black} Consider a general gradient flow with mobility, which is described by the PDE
\begin{equation*}
  \partial_t h=-\A_h\left(\frac{\delta E}{\delta h}\right).
\end{equation*}
This form includes the following standard models:
\begin{itemize}
\item
Allen-Cahn:  $\A_h = \mbox{I}$ (unit operator); and \color{black}
\item 
Cahn-Hilliard:  $\A_h=-\Delta$.
\end{itemize}
Here, for the energy functional we take
$$
E[h]=\int |\nabla h|^p \ud x~,
$$
which is convex ($p\ge 1$). \color{black}

For typical diffusion-limited kinetics in crystal relaxation~\cite{GigaKohn}, we have
\[
\A_h=-\nabla \cdot (M(|\nabla h|)\nabla)~,
\]
with mobility $M(|\nabla h|)$. Thus, the PDE reads 
\begin{equation*}
\partial_t h=\nabla\cdot\left(M(|\nabla h| ) \nabla \big(\frac{\delta E}{\delta h}\big) \right)~.
\end{equation*}
Let us now discretize in time so that $h^n:=h(x,n\tau)$, where $\tau$ is the timestep, and define the function \color{black}
\begin{equation*}
   \Phi(\tau,h^{n-1}; h):=E(h)+\frac{1}{2\tau}{\textbf{dist}}^2(h, h^{n-1})~,
\end{equation*}
where
\begin{equation*}
 {\textbf{dist}}^2(h,h^{n-1})=\int (h-h^{n-1}) \A_{h^{n-1}}^{-1}(h-h^{n-1}) \ud x~.
\end{equation*}

We then freeze $h$, hence freezing the mobility $M(|\nabla h|)$, and regard $\textbf{dist}$ 
as a metric for the (weighted) $H^{-1}$ metric space.
We view the subdifferential for the convex functional $E(h)$ in this frozen metric space as the actual subdifferential of our theory in this paper. We can then discretize in time by using the unconditionally-stable backward Euler scheme~\cite{AGS}, viz., \color{black}
\begin{equation*}
h^{n} \in {\textnormal{argmin}} \Phi(\tau,h^{n-1}; \cdot)~.
\end{equation*}
A well-known example for this machinery can be found in~\cite{JKO}, where
$\A_h=-\nabla \cdot ( h \nabla)$, ${\textnormal{dist}}(h,h^{n-1})=W_2(h,h^{n-1})$ and $W_2$ is the Wasserstein cost functional. In the present paper, we extend this machinery to the operator $\A_h$ corresponding to exponential PDE~\eqref{eq:PDE-evol}. \color{black}
 
%%%%%%%%%%%%%%%%%%%%%%%%%%%%%%%%%%%%%%%%%%%%%%%%%%%%%%%%%%%%%%%%%%%%%%%%%
\subsection{ODE dynamics from subdifferential formulation}
\label{subsec:ode-dynamics}
%%%%%%%%%%%%%%%%%%%%%%%%%%%%%%%%%%%%%%%%%%%%%%%%%%%%%%%%%%%%%%%%%%%%%%%%%
Based on the above formalism, we proceed to derive ODEs for facet motion in the exponential total variation flow
\begin{equation}
\label{exppde}
\partial_t h = \partial_{xx} e^{- \partial_x \left( \frac{\partial_x h}{ | \partial_x h |}\right)}, \ \ h(x,0) = h_0 (x)~,
\end{equation}
using the natural profile stemming from a regularized solution for the $1$-Laplacian.

Evidently, PDE~\eqref{exppde} has the structure
\begin{equation}\notag
\begin{aligned}
& \partial_t h = - \partial_x J \quad (\text{continuity equation})~, \\ 
& J = - \partial_x \varrho \quad \quad (\text{Fick's law})~, \\
& \varrho = e^\mu \quad \quad (\text{Gibbs-Thomson relation})~,\\
& \mu = \frac{ \delta E}{ \delta h} =  - \partial_x w(\partial_x h) \quad (\text{thermodynamic force})~,
\end{aligned}
\end{equation} 
where $J$ is the (scalar) mass flux.

We spell out the following simplifying symmetry assumptions (a few of which we have already mentioned above):
\begin{itemize}
\item The facet solution is symmetric (with respect to $x=0$), i.e., $h(-x,t) = h(x,t)$.
\item The facet has zero slope, i.e., $\partial_x h  = 0$ for $x \in (-x_f,x_f)$.  In addition, for a top facet we have $\partial_x h < 0$ when $x > x_f$, and for a bottom facet we have $\partial_x h > 0$ when $x > x_f$.
\item The function $w(u) = u/|u|$ ($u$ is the slope) is extended onto the facet as an odd function on $\RR$. We set $\tilde w (x,t) = w(\partial_x h)$.  \color{black}
\item The mass flux $J(x,t)$ is an odd function on $\mathbb{R}$, i.e., $J(-x,t)=-J(x,t)$. 
\color{black}
\end{itemize}

On the facet, where $0 < x < x_f (t)$ and $h(x,t) = h_f (t)$, we therefore obtain
\begin{equation}\notag
\dot{h}_f = - \partial_x J~,
\end{equation}
which by integration yields
\begin{equation}\notag
J(x,t) = -x \dot{h}_f + C_1 (t)~.
\end{equation}
Note that $C_1 (t) = 0$ by the symmetry considerations that $h$ will be even.    In addition, as in the total variation flow observations of \cite{GigaGiga}, we are compelled to recognize a jump in $h$ at $x = x_f (t)$, forcing the remaining functions $J$ and $\mu$ to have continuous derivatives.  

By the PDE structure, we additionally have
\begin{equation}\notag
\partial_x (e^\mu) = -J(x,t)~,
\end{equation}
which entails
\begin{equation}\notag
\mu(x,t) = \ln \left(  \frac{x^2}{2} \dot{h}_f + C_2 (t) \right).
\end{equation}
We also have
\begin{equation}\notag
\partial_x \tilde w = - \ln \left(  \frac{x^2}{2} \dot{h}_f + C_2 (t) \right),
\end{equation}
which is integrated to give
\begin{equation}\notag
\tilde w(x,t) = - \int_0^x  \ln \left(  \frac{s^2}{2} \dot{h}_f + C_2 (t) \right) \dv s + C_3 (t)~.
\end{equation}
The integration constant $C_3(t)$ is determined by our assumption that $\tilde w(x,t)$ is an odd function of $x$; hence, $C_3 = 0$. \color{black} 

In addition, mass conservation dictates that 
\[
\int_0^{x_f} h_0 (s) ds = h_f x_f~,
\]
which yields the motion law
\begin{equation}\label{eq:mass-cons1}
\dot{x}_f (h_0 (x_f) - h_f) = \dot{h}_f x_f~.
\end{equation} 

\subsection{Dynamics of top facet}

At this stage, we need to specify if the symmetric facet lies in the convex or concave part of the surface. This choice affects the sign of $\dot h_f$ and leads to different dynamics, as we show below.
Let us begin with the case in which the facet is a degenerate local maximum.  

By continuity of $w$ and $\mu$, the following conditions hold:
\begin{equation}\notag
\begin{aligned}
& \tilde w (x_f,t) = - \int_0^{x_f} \ln \left(  \frac{s^2}{2} \dot{h}_f + C_2 (t) \right) \dv s = -1~, \\
& \mu (x_f,t) =  \ln\left(  \frac{x_f^2}{2} \dot{h}_f + C_2 (t) \right) = 0~,
\end{aligned}
\end{equation}
where
\begin{equation}\notag
	\dot h_f <0~.
\end{equation}
The above condition for $\mu$ yields
\begin{equation}\notag
C_2 (t) = 1 - \frac{x_f^2}{2} \dot{h}_f\quad (C_2>0)~.
\end{equation}
Here, $\tilde w(x_f,t) = -1$, since $x=x_f$ lies at the right endpoint of the top facet and away from the maximum. (Note that $w(h_x )=h_x/|h_x|= -1$ on the right of the facet).\color{black}

Hence, we obtain the system
\begin{equation}
\left\{  \begin{array}{c}
 \int_0^{x_f}  \ln \left(  \frac{s^2}{2} \dot{h}_f + 1 - \frac{x_f^2}{2} \dot{h}_f \right) \dv s = 1~,   \\
 \dot{x}_f (h_0 (x_f) - h_f) = \dot{h}_f x_f~.
\end{array} \right.  \label{facetode}
\end{equation}
The first equation reads \color{black}
\begin{equation}\notag
\int_0^{x_f \sqrt{ \frac{| \dot{h}_f |}{2}}  }   \ln \left(   1 - \frac{x_f^2}{2} \dot{h}_f - \xi^2 \right) \dv\xi = \sqrt{ \frac{| \dot{h}_f |}{2}}~.
\end{equation}
By using the integral
\[
\int_0^a   \ln \left(   1 + a^2-  \xi^2 \right) \dv \xi = -2 a + 2 i \sqrt{1 + a^2} \tan^{-1} \left( -i \frac{a}{\sqrt{1 + a^2}} \right)
\]
along with the definition
\begin{equation}\notag
X_f = x_f  \sqrt{ \frac{| \dot{h}_f |}{2}}~,
\end{equation}
we arrive at the system 
\begin{equation}\notag
\left\{  \begin{array}{c}
2  \sqrt{1 + X_f^2}  \ln \left( X_f + \sqrt{ 1 + X_f^2}   \right) - 2 X_f =  \sqrt{ \frac{| \dot{h}_f |}{2}}~, \\
 \dot{x}_f (h_0 (x_f) - h_f) = \dot{h}_f x_f~.
 \end{array}  \right.
\end{equation}

This is a closed ODE system describing the top-facet dynamics.  Because $\dot h_f < 0$, we may frame the system of equations as a system of differential-algebraic equations (DAE), viz.,
\begin{equation}\notag
\begin{aligned}
& X_f = x_f  \sqrt{ \frac{| \dot{h}_f |}{2}}~, \\
& 2 \sqrt{1 + X_f^2}   \ln \left( X_f + \sqrt{ 1 + X_f^2}   \right)   - 2 X_f =  \sqrt{ \frac{| \dot{h}_f |}{2}}~, \\
&  \dot{x}_f (h_0 (x_f) - h_f) = \dot{h}_f x_f~.
\end{aligned}
\end{equation}
These equations are now recast into a form that can be solved by implicit ODE solvers, viz., \color{black}
\begin{equation}
\left\{ \begin{array}{c}
 \dot{X}_f = \frac{ \dot{x_f} F(X_f)}{ 1- x_f F'( X_f)}~, \\
 \dot{h}_f = -2 F(X_f)^2~,  \\
 \dot{x}_f (h_0 (x_f) - h_f) = -2 x_f F(X_f)^2~,
 \end{array} \right.    \label{daesys}
\end{equation}
where
\[
F(X_f) = 2 \sqrt{1 + X_f^2} \ln( X_f + \sqrt{1 + X_f^2}) - 2 X_f~.
\]
It is of interest to note that the algebraic equation for $X_f$ suggests that the correct value for $X_f (0)$ is given by a solution of
\begin{equation}
\label{Xfeq}
  x_f (0) \left[2 \sqrt{1 + X_f^2}   \ln \left( X_f + \sqrt{ 1 + X_f^2}   \right)   - 2 X_f \right] - X_f = 0~,
\end{equation}
which has three roots given by $X_f = 0,\, \pm g(x_f)$.  The non-zero roots $\pm g(x_f)$ for large $X_f$ should take the form
\[
g(x_f) \approx \frac{e^{\frac{1}{2 x_f (0)} +1}}{2}~.
\]

We reach the conclusion that, under the dynamics of \eqref{daesys}, there is {\it no restriction} on the initial width, $2 x_f(0)$, of the facet for the expansion of the facet at later times, $t > 0$.  

\subsection{Dynamics of bottom facet}

Let us now discuss the case where the facet possibly corresponds to a degenerate local minimum of the height profile. In this case, we assume to have 
\begin{equation}\notag
\dot h_f > 0~.
\end{equation}
The dynamics in \eqref{facetode} are replaced by the system
\begin{equation}
\left\{  \begin{array}{c}
 \int_0^{x_f}  \ln\left(  \frac{s^2}{2} \dot{h}_f + 1 - \frac{x_f^2}{2} \dot{h}_f \right) \dv s = -1~,   \\
 \dot{x}_f (h_0 (x_f) - h_f) = \dot{h}_f x_f~.
 \end{array} \right.  \label{facetode_bottom}
\end{equation}  
The first equation implies that $\frac{x_f^2}{2 \dot{h}_f }\leq 1$.  

Accordingly, by changing variables we observe that
\begin{equation}
\label{int_min}
\int_0^{X_f  }   \ln \left(    \xi^2 +1 - X_f^2 \right) \dv\xi =- \sqrt{ \frac{\dot{h}_f }{2}} \leq 0~.
\end{equation}
By integrating directly in view of $\dot{h}_f > 0$, we obtain the system
\begin{equation}\notag
\begin{aligned}
& 2  \sqrt{1 -X_f^2}  \left[ \tan^{-1} \left( \frac{   \sqrt{1 -X_f^2}  }{ X_f} \right) - \frac{\pi}{2}    \right] + 2 X_f =  \sqrt{ \frac{ \dot{h}_f }{2}}~, \\
&  \dot{x}_f (h_0 (x_f) - h_f) = \dot{h}_f x_f~.
\end{aligned}
\end{equation}
The first equation can be written as
\begin{equation}
\label{yeq}
y \left( \tan^{-1} y - \frac{\pi}{2} \right) + 1 = \frac{1}{2 x_f}
\end{equation}
where
\[
y = \frac{\sqrt{ 1 - X_f^2}}{X_f}~.
\]
The left-hand side of \eqref{yeq} is bounded by $1$ for $y \geq 0$ (in fact, it is monotonically decreasing from $1$); while, if  $x_f$ is sufficiently small, the right-hand side gets arbitrarily large.  Hence, we reach the conclusion that: {\em only if $x_f (0) > \frac12$ is it possible to find a solution with a moving bottom facet.}

As a result, facet solutions at minima are in fact fixed points of the evolution unless there is already a sufficiently long facet.  This asymmetry in convexity and concavity of the morphological crystal surface evolution is consistent with observations of solutions to the exponential PDE in~\cite{MW1}.  

%Also, the time scales of ODEs for \eqref{facetode_bottom} is comparable to that of the relaxation time scale for a regularization parameter we were able to reach and we could not compare cases that did not move to those that do.  

\medskip

\noindent {\bf Remark 1.}  For the dynamics given by \eqref{eq:PDE-evol} as a nonlinear (weighted) $H^{-1}$ flow, \color{black} the evolutions of top and bottom facets are distinctly different, precisely because of the effect of the exponential mobility, $e^{-\partial_x (\partial_x h/|\partial_x h|)}$.  In particular, we predict that bottom facets have extremely slow (or non-existent) motion by diffusion, while top facets move relatively fast.

%%%%%%%%%%%%%%%%%%%%%%%%%%%%%%%%%%%%%%%%%%%%%%%%%%%%%%
%%%%%%%%%%%%%%%%%%%%%%%%%%%%%%%%%%%%%%%%%%%%%%%%%%%%%%
\section{Numerical results}
\label{sec:numerics}
%%%%%%%%%%%%%%%%%%%%%%%%%%%%%%%%%%%%%%%%%%%%%%%%%%%%%%
%%%%%%%%%%%%%%%%%%%%%%%%%%%%%%%%%%%%%%%%%%%%%%%%%%%%%%
In this section, we present numerical results for the evolution of the height profile under sinusoidal initial data in 1+1 dimensions. We carry out numerics based on: (i) the ODE system discussed in Section~\ref{sec:ODE-facets}; and (ii) the numerical solution of PDE~\eqref{eq:PDE-evol} via the regularization of free energy~\eqref{eq:cont-surf-en}.
Specifically, we use the regularized  surface free energy 
 \begin{equation}
\label{ereg}
E [h;\nu] = \int \left[\sqrt{ | \nabla h|^2 + \nu^2} + \frac{g}{3} |\nabla h|^3 \right] \dv x~,
\end{equation}
which has the regularization parameter $\nu>0$.

\subsection{Numerical approximation with $g=0$}
Next, we focus on the regularized versions of PDEs \eqref{exppde} and \eqref{gigapde}. The corresponding PDEs now read
\begin{equation}
\label{exppde_visc}
\partial_t h = \partial_{xx} \exp\left\{- \partial_x  \left( \dfrac{  \partial_x h}{  \sqrt{  (\partial_x h)^2 + \nu^2} }  \right)\right\}
\end{equation}
and
\begin{equation}
\label{gigapde_visc}
\partial_t h = -\partial_{xx}  \partial_x  \left(  \frac{  \partial_x h}{  \sqrt{  (\partial_x h)^2 + \nu^2} } \right)~.
\end{equation}
For discretizing both \eqref{exppde_visc} and \eqref{gigapde_visc}, we apply a standard central finite difference discretization in space with a fully implicit stepping scheme in time (by using routine \textsf{ode15s} in \textsf{MATLAB}).

\begin{figure}[t]
\begin{center}
\includegraphics[width=.45\textwidth]{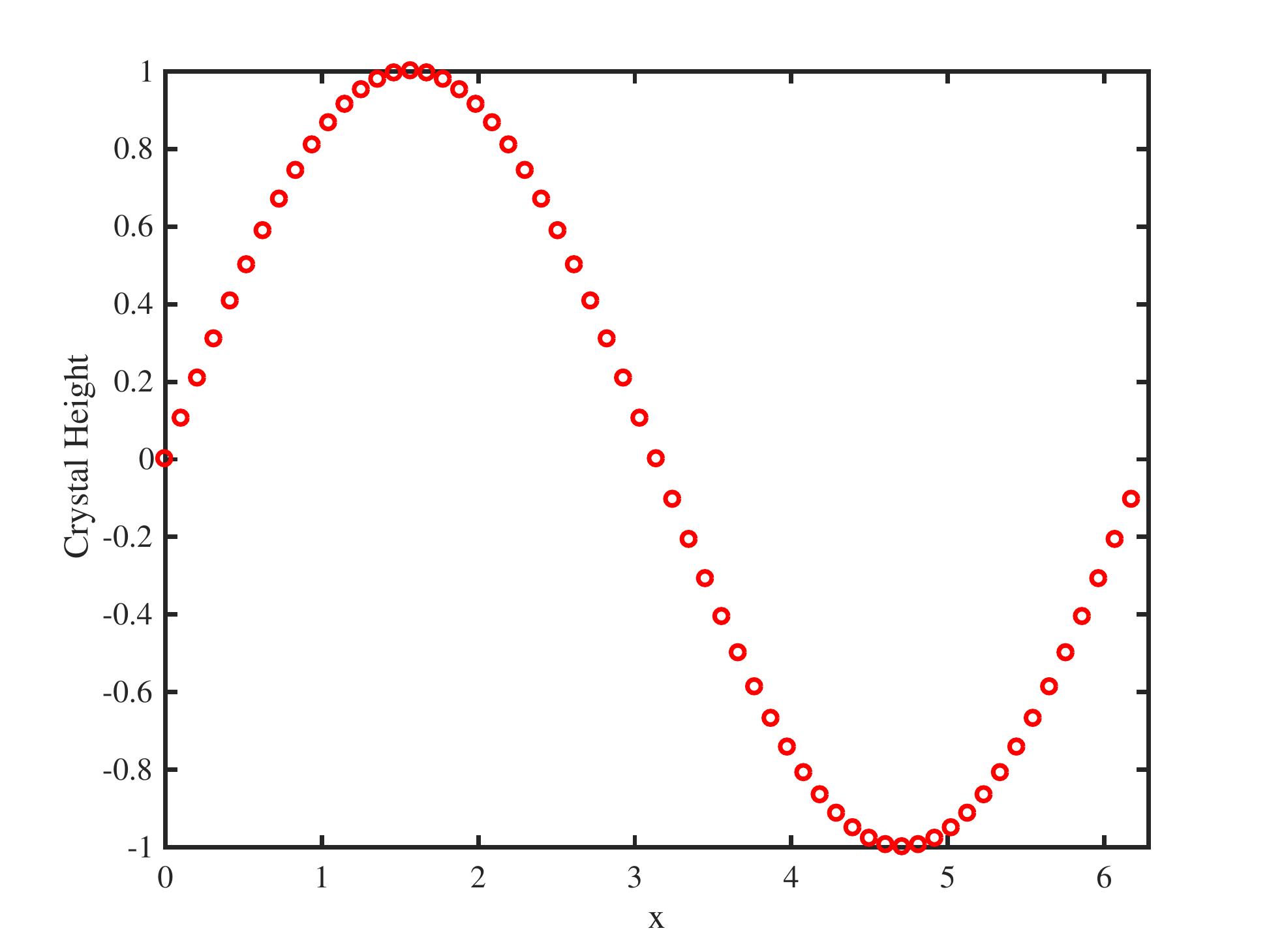} \\
\includegraphics[width=.45\textwidth]{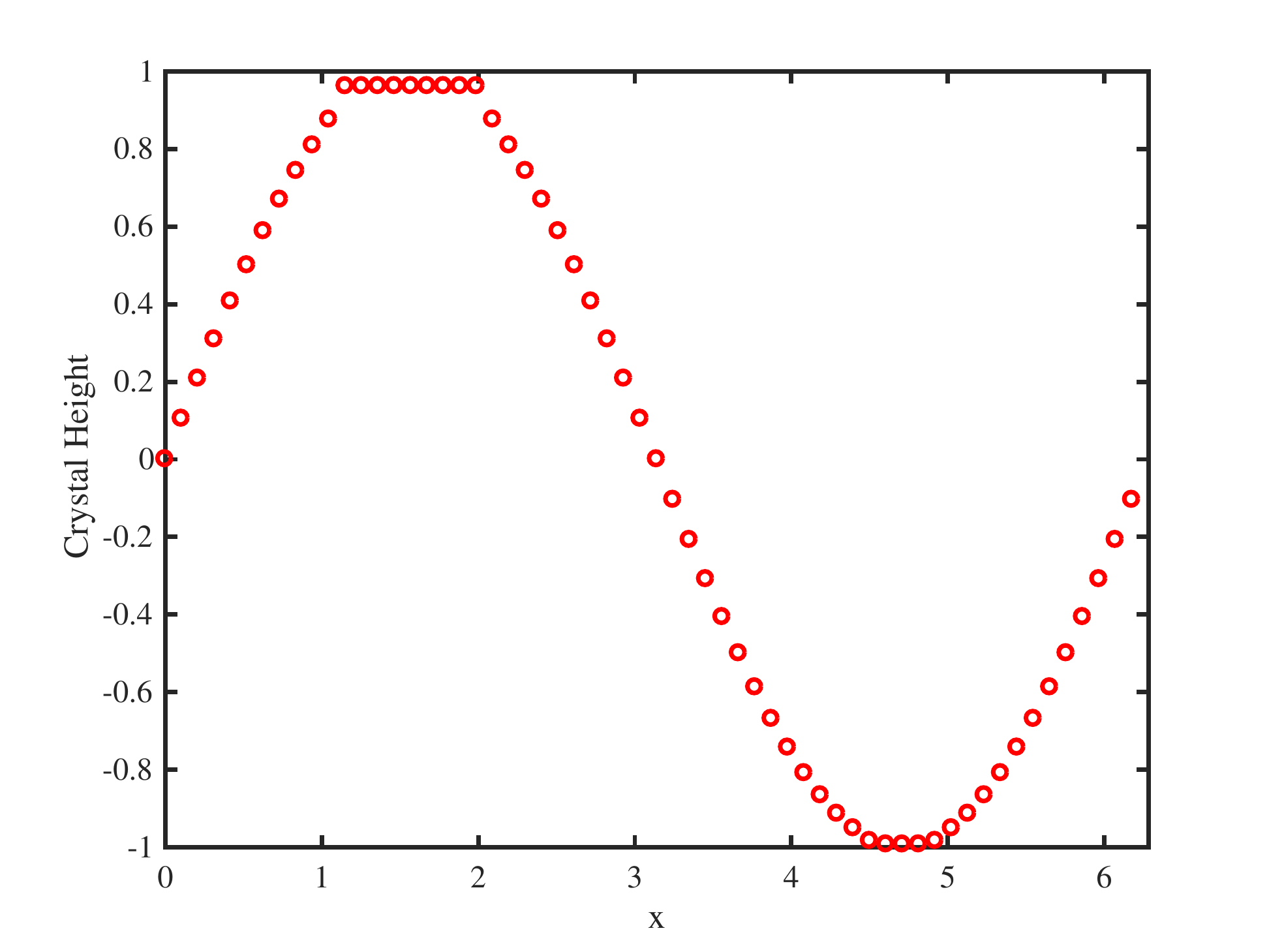}  
\includegraphics[width=.45\textwidth]{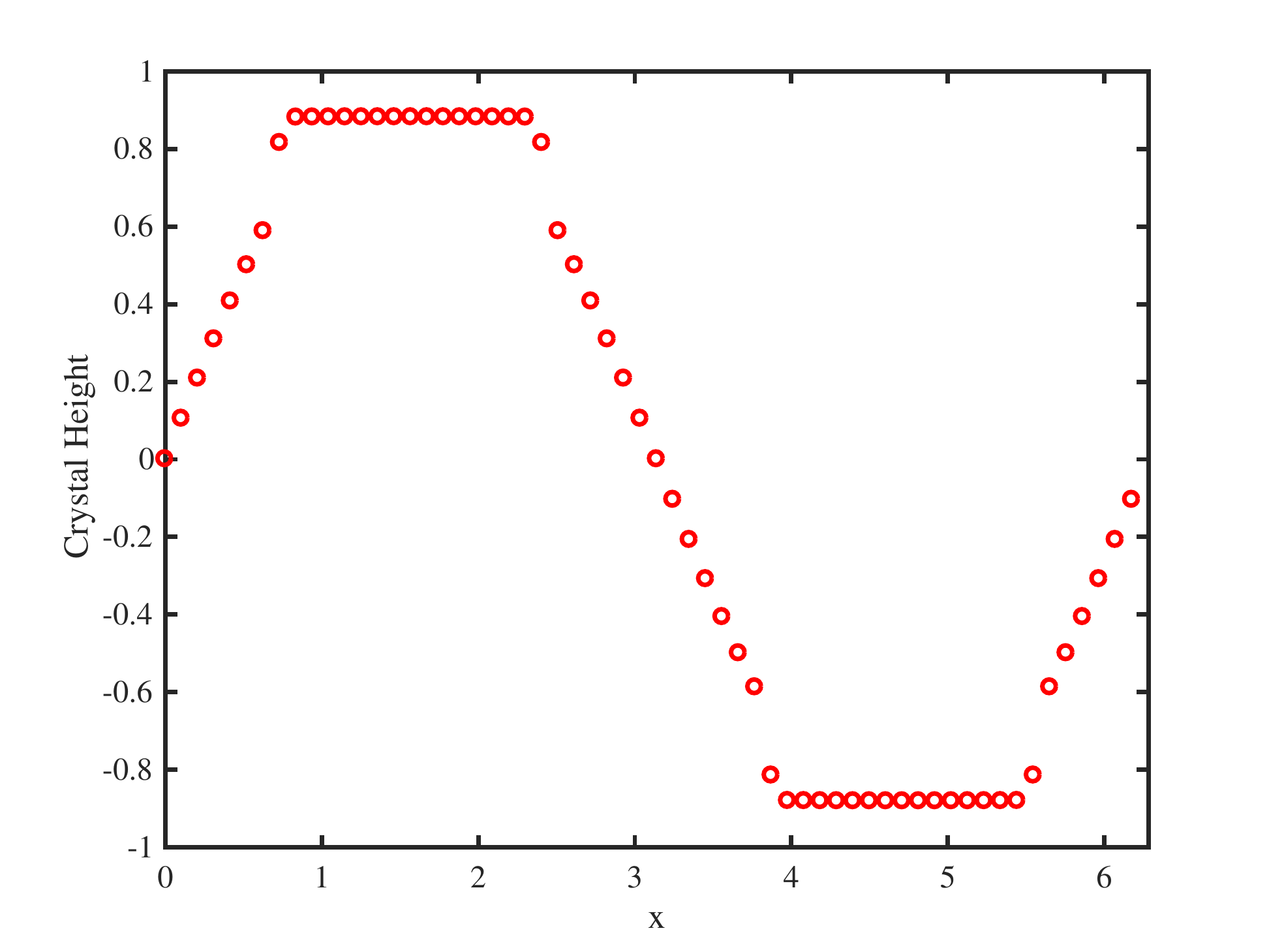}  
\caption{Snapshots of evolving surface height profile, $h(x,t)$,  under initial data $h (x,0) = \sin ( x)$ (top panel) by fourth-order total variation flows given by: exponential PDE~\eqref{exppde_visc} with regularization parameter $\nu =  10^{-3}$ on a time scale $T = 10^{-4}$ (bottom left panel);  and by PDE~\eqref{gigapde_visc} with regularization parameter $\nu = 10^{-3}$ on a time scale $T = 10^{-2}$ (bottom right panel).  }
\label{f:shocks}
\end{center}
\end{figure}

Snapshots of solutions to evolution equations \eqref{exppde_visc}  and  \eqref{gigapde_visc} under an initial height profile 
$h (0,x) = \sin x$ with $N = 60$ uniform grid points on the interval $[0,2 \pi]$ and $\nu = 10^{-3}$ by use of periodic boundary conditions can be seen in Figure \ref{f:shocks}.  We have chosen time scales such that the facets are evident in the numerical solutions. Note that exponential PDE~\eqref{exppde_visc} results in a strong asymmetry between regions of convexity and concavity, seen in the bottom left panel of Figure~\ref{f:shocks}. \color{black}  For each simulation, we have chosen the regularization parameter and the grid spacing such that the resulting derivatives are sufficient to allow facet motion but also to maintain a sharp facet boundary. In contrast to the clear convex/concave asymmetry of the numerical solution to \eqref{exppde_visc}, notice the symmetry in the solution of~\eqref{gigapde_visc} in the bottom right panel of Figure~\ref{f:shocks}. \color{black}

\begin{figure}[t]
\begin{center}
\includegraphics[width=.45\textwidth]{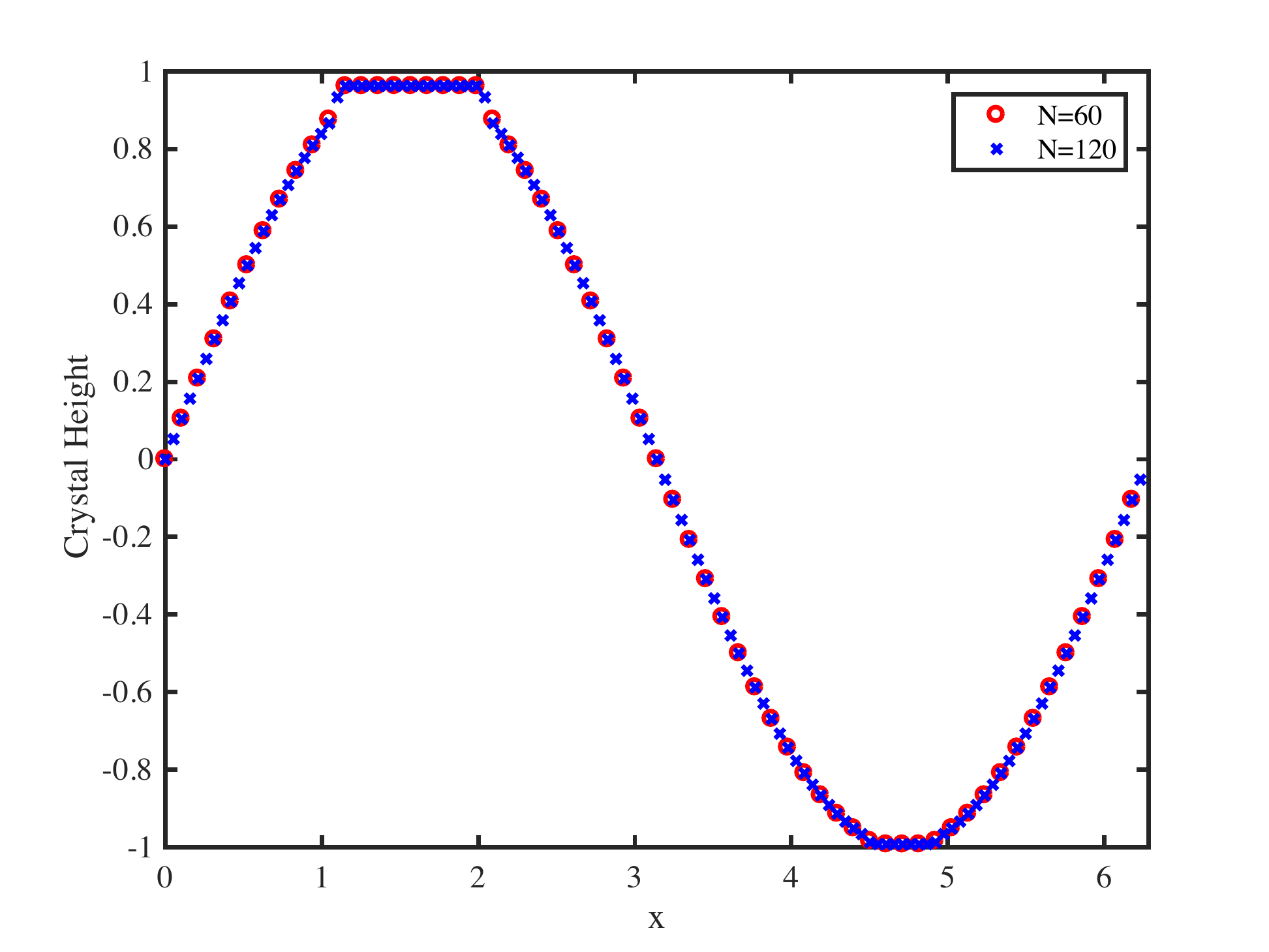}
\caption{(Color Online) Snapshots of evolving surface height profile, $h(x,t)$,  for two different values ($N=60,\,120$) of the number, $N$, of grid points on a time scale $T=10^{-4}$. The height $h(x,t)$ evolves according to exponential PDE~\eqref{exppde_visc} with initial data $h (x,0) = \sin ( x)$ and regularization parameter $\nu =  10^{-3}$.  }
\label{fig:numcomp}
\end{center}
\end{figure}

To verify that our numerical scheme is consistent, that is, the increase of resolution (number of grid points) improves or does not spoil the result, we plot
a snapshot of the solution to \eqref{exppde_visc} at time $T = 10^{-4}$ with $N=60$ and $N=120$ grid points; see Figure~\ref{fig:numcomp}. Notice that the numerical solution is stabilized, remaining practically unchanged with  increasing $N$. \color{black}

\begin{figure}[t]
\begin{center}
\includegraphics[width=.45\textwidth]{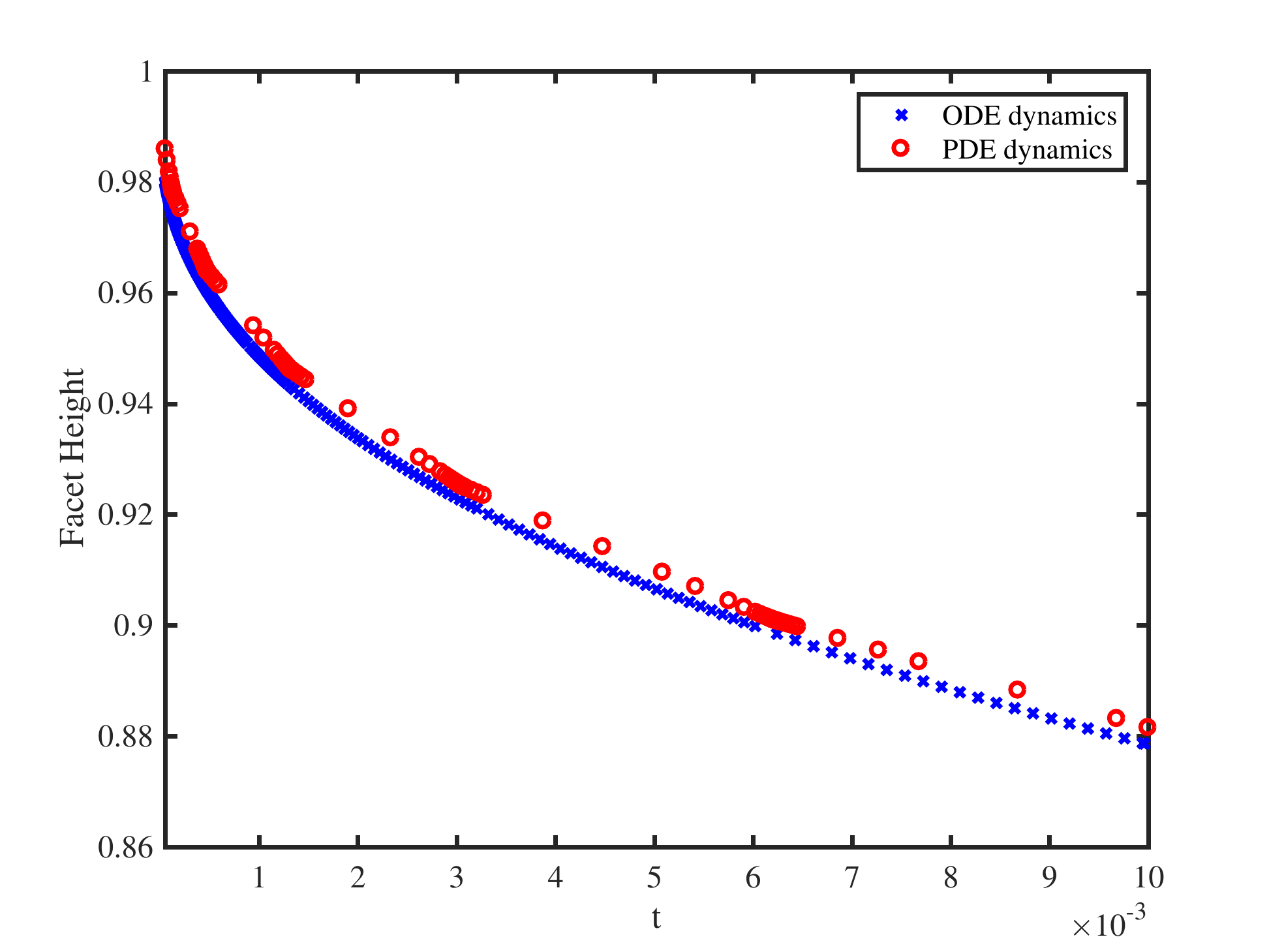}
\includegraphics[width=.45\textwidth]{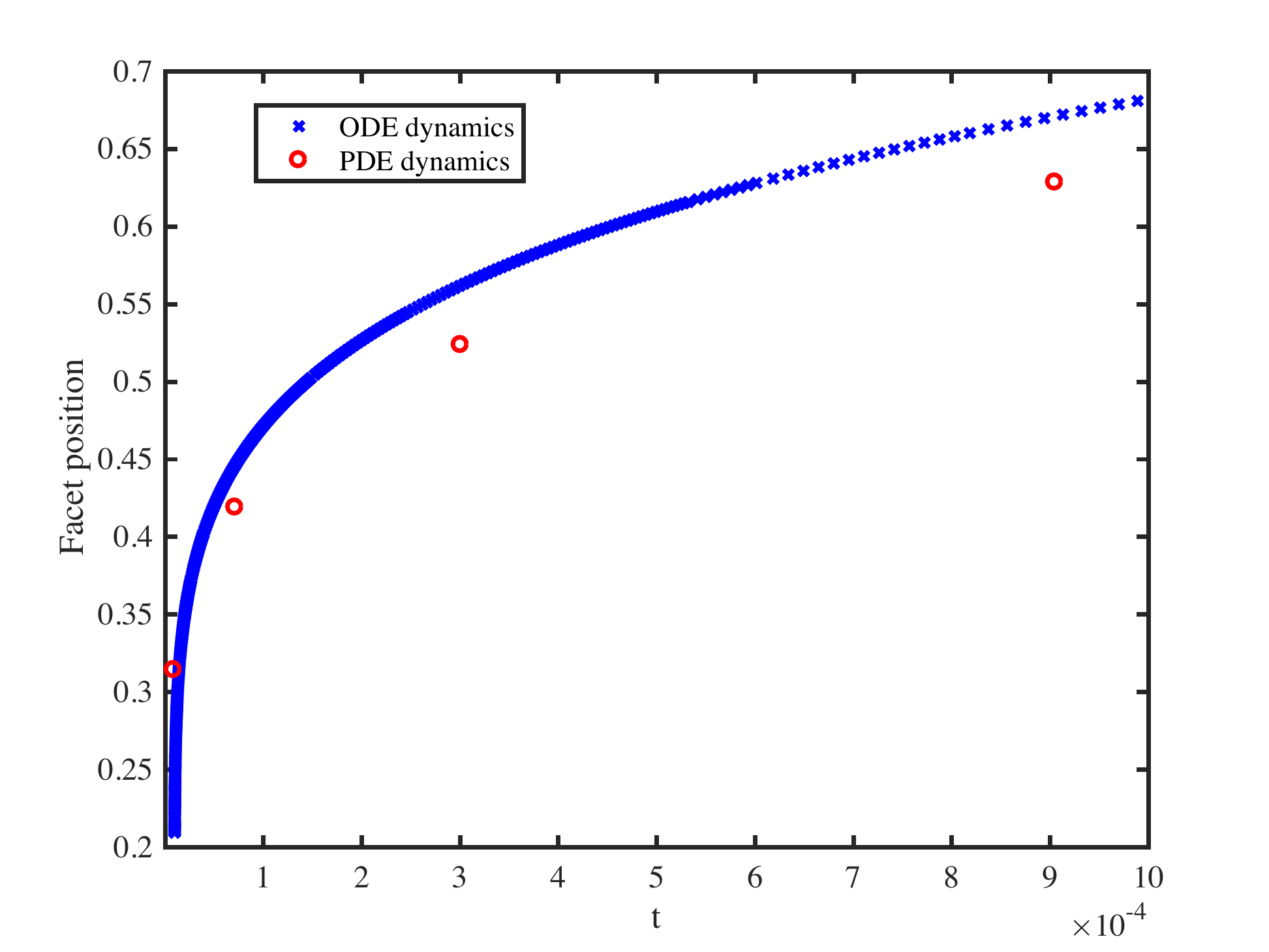} \\
\includegraphics[width=.45\textwidth]{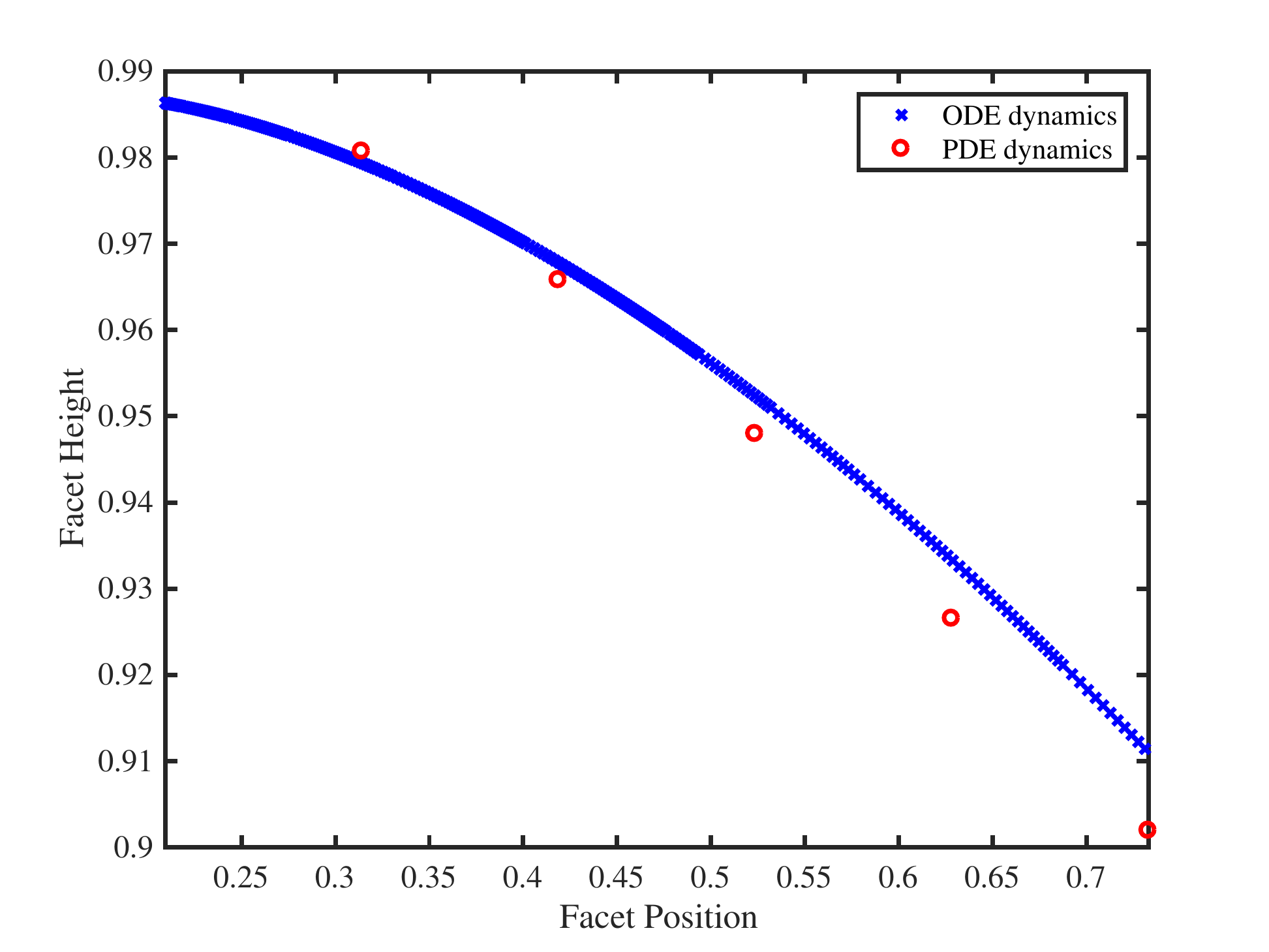}
\caption{(Color Online)  Plots of facet height $h_f (t)$ versus time, $t$ (top left panel), facet position $x_f(t)$ versus $t$ (top right panel) and facet height versus facet position ($x_f (t)$, $h_f (t)$) (bottom panel) for exponential PDE~\eqref{exppde_visc}.   In each plot, $(x_f,h_f)$ as a solution of \eqref{daesys} is plotted using crosses ($\times$); and the corresponding components of a solution to \eqref{exppde_visc} as described in \eqref{pdeproj} are plotted using circles ($\circ$).  The initial data for \eqref{daesys} is taken from the PDE evolution as $x_f (t_0) = \frac{\pi}{15}$, $h_f(t_0) = .98879899$ with $t_0 = 5 \times 10^{-7}$.  The numerical experiments for the ODEs and PDE are then compared up to time $T = 10^{-3}$.}
\label{f:compexp}
\end{center}
\end{figure}

 In Figure \ref{f:compexp}, the evolution of PDE~\eqref{exppde_visc} is compared to ODE system~\eqref{daesys} on time scales such that the facets are evident.  In these simulations, the parameters for the PDE simulation are the same as above.  To solve DAEs~\eqref{daesys}, we use the implicit ODE solver~\textsf{ode15i} in \textsf{MATLAB} with explicitly chosen initial data for $X_f (0)$ as a non-zero root of~\eqref{Xfeq}. To generate the initial data $(x_f (0), h_f(0))$, we find it ideal to numerically solve PDE~\eqref{exppde_visc} for a short time ($\approx 5\times10^{-7}$).  Then, we generate a non-singular  (i.e. with $x_f (0) > 0$) initial configuration for the ODEs by reading off the maximum height of the resulting numerical facet solution and the outer extent of the facet position.  Note that there is some sensitivity in how the initial data $x_f (0)$ is chosen given the discretization, which explains the small discrepancy observed in those plots involving $x_f(t)$.  To compare the relevant parameters to the PDE evolution, we take 
 \begin{align}
\label{pdeproj}
 h_{f,pde} (t) &= \max_{x \in [0,2 \pi]} h(x,t)~,\notag\\
 x_f&= \max \{ x \in [0, 2\pi] :  (\max_{x \in [0, 2\pi]} h(x,t)) - h(x,t) < \varepsilon \}~,
 \end{align}
 where we typically choose $\varepsilon = 10^{-2}$.  The data points for $x_f (t)$ in Figure \ref{f:compexp} appear to occur on larger time scales than the discretization would suggest.  However, this is purely a manifestation of the time required for the facet edge to travel from one discrete grid point to another in the numerical experiment.  To make the figure clearer, we have thus only plotted times at which the solution has moved to a new grid point; the large gaps in data points for $x_f$ are due entirely to the spatial grid size. 

% \jl{Is it possible to do some sub-grid resolution postprocessing trick (say for example, interpolate the profile in a finer grid using some filters)?}
%  \jlm{How do you like the solution I have attempted??}
 
\begin{figure}[t]
\begin{center}
\includegraphics[width=.45\textwidth]{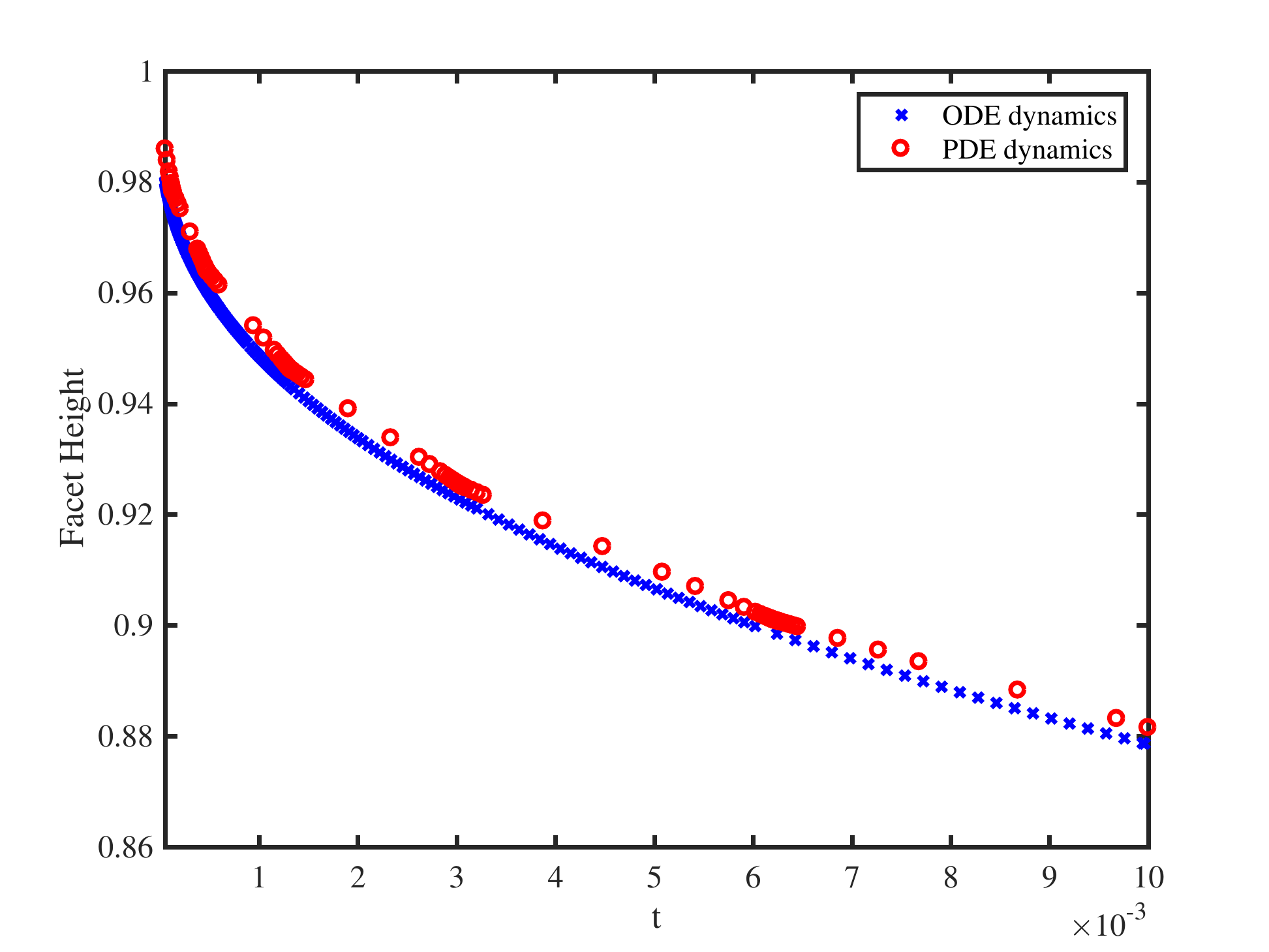}
\includegraphics[width=.45\textwidth]{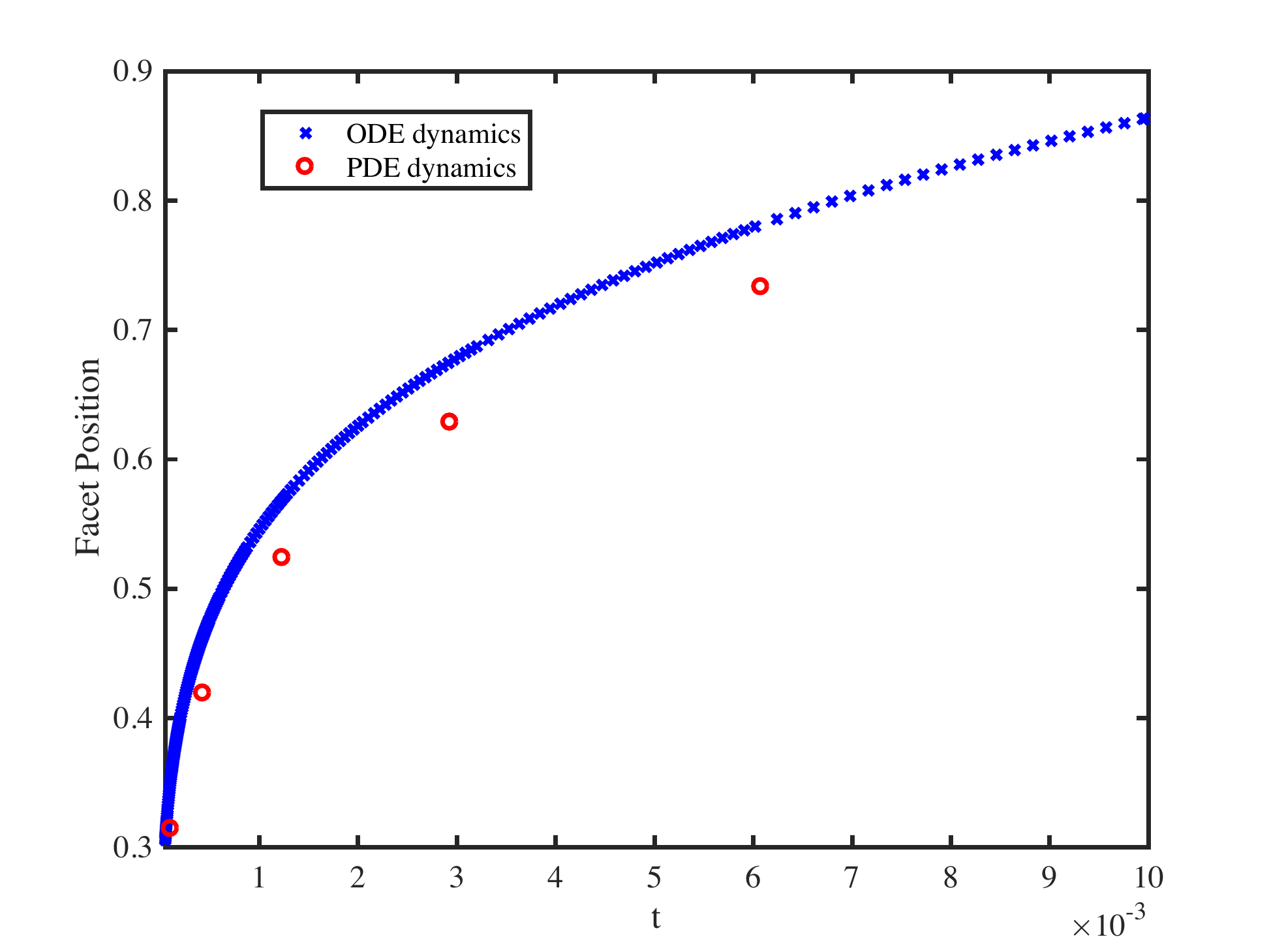} \\
\includegraphics[width=.45\textwidth]{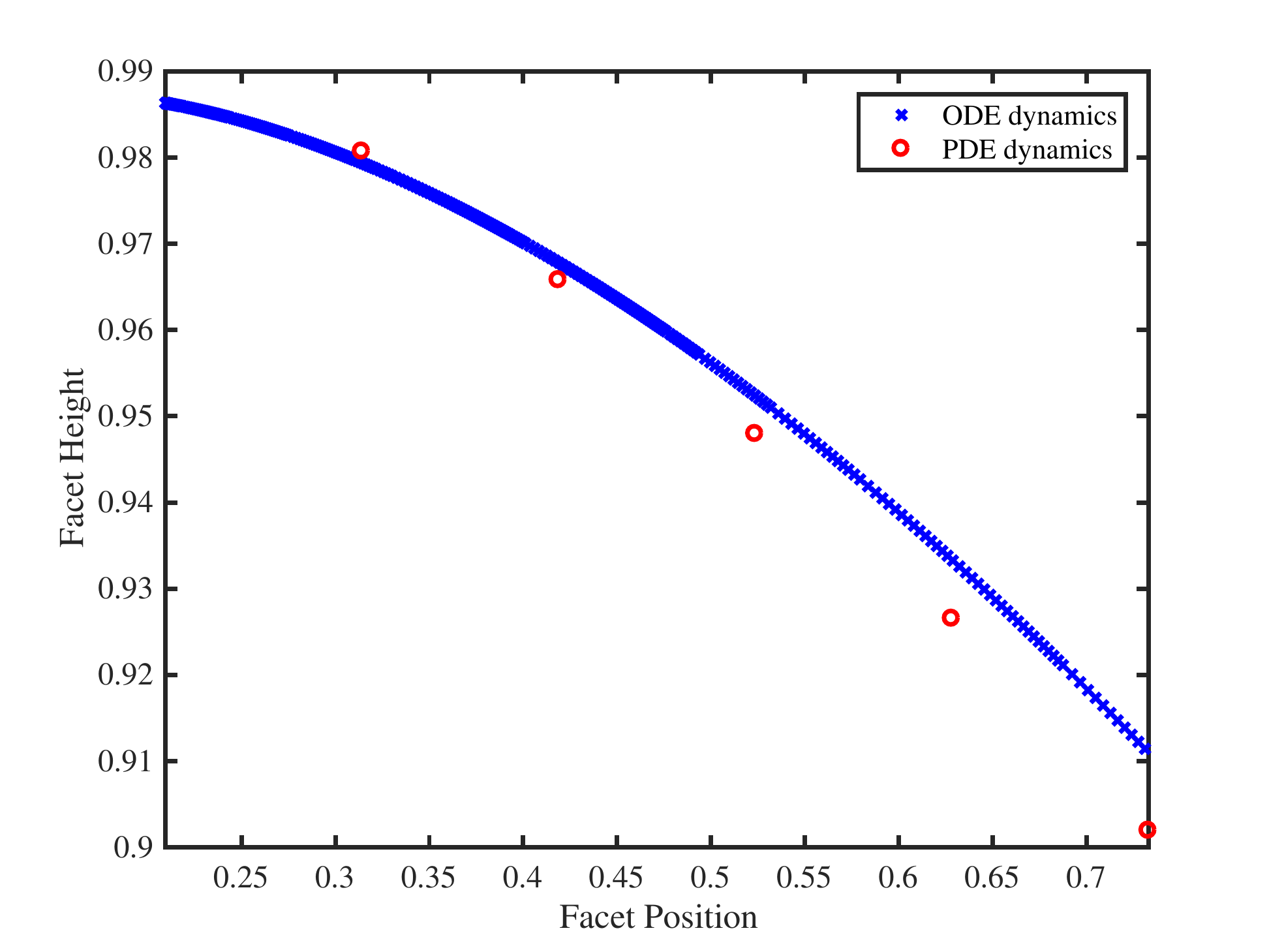}
\caption{(Color Online)  Plots of facet height $h_f (t)$ versus time, $t$ (top left panel), facet position $x_f(t)$ versus $t$ (top right panel) and facet height versus facet position ($x_f (t)$, $h_f (t)$) (bottom panel) for PDE~\eqref{gigapde_visc}.   In each plot, $(x_f,h_f)$ as a solution of \eqref{gigasys} is plotted using crosses ($\times$); and the corresponding components of a solution to \eqref{gigapde_visc} as described in \eqref{pdeproj} are plotted using circles ($\circ$).  The initial data for \eqref{gigasys} is taken from the PDE evolution as $x_f (t_0) = \frac{\pi}{15}$, $h_f(t_0) = .98632751$ with $t_0 = 5 \times 10^{-5}$.  The numerical experiments for the ODEs and PDE are then compared up to time $T = 10^{-2}$.}
\label{f:compGiga}
\end{center}
\end{figure}

  In Figure \ref{f:compGiga}, we carry out a similar numerical study as in Figure \ref{f:compexp}, but now for the (non-weighted) $H^{-1}$ total variation flow \eqref{gigapde_visc} studied, e.g., in \cite{KobayashiGiga99,GigaGiga,GigaKohn,Giga1}.  The PDE evolution is compared to the ODE system \eqref{gigasys} on time scales such that the (top and bottom) facets are evident.  In these simulations, the discretizations for the PDE are the same as those used for the exponential PDE in this section.

%{\color{blue}  Discuss difficulties with JKO/Implicit Time stepping scheme of \cite{KohnV} here?}

\subsection{Numerical approximation with $g>0$}

In this subsection, we focus on the case with nonzero step-step interactions ($g > 0$); see \eqref{eq:cont-surf-en} and~\eqref{ereg}.  Accordingly, we consider the fourth-order PDE
\begin{equation}
\label{exppde_reg}
\partial_t h = \partial_{xx} e^{- \partial_x \left( \frac{\partial_x h}{ | \partial_x h |} + g \partial_x h |\partial_x h| \right)}~, \quad g >0~.
\end{equation}
  In this setting, we still observe asymmetry in the solution.  However, due to presence of the (less singular) term $|\partial_x h|^3$ in the surface energy, the solution to this PDE no longer develops jumps in the height profile.  This is expected from other studies in the (non-weighted) $H^{-1}$ total variation flow; see e.g. \cite{KohnV}.  Similarly to the case where $g=0$, we can study the evolution numerically by using the regularized flow
  \begin{equation}
\label{exppde_visc_reg}
\partial_t h = \partial_{xx} e^{ - \partial_x  \left( \frac{  \partial_x h}{  \sqrt{  (\partial_x h)^2 + \nu^2} }  +g  \partial_x h |\partial_x h|  \right)}~,
\end{equation}
which corresponds to free energy $E[h;\nu]$ of~\eqref{ereg}.

\begin{figure}[t]
\begin{center}
\includegraphics[width=.45\textwidth]{Fig2a.pdf}
\includegraphics[width=.45\textwidth]{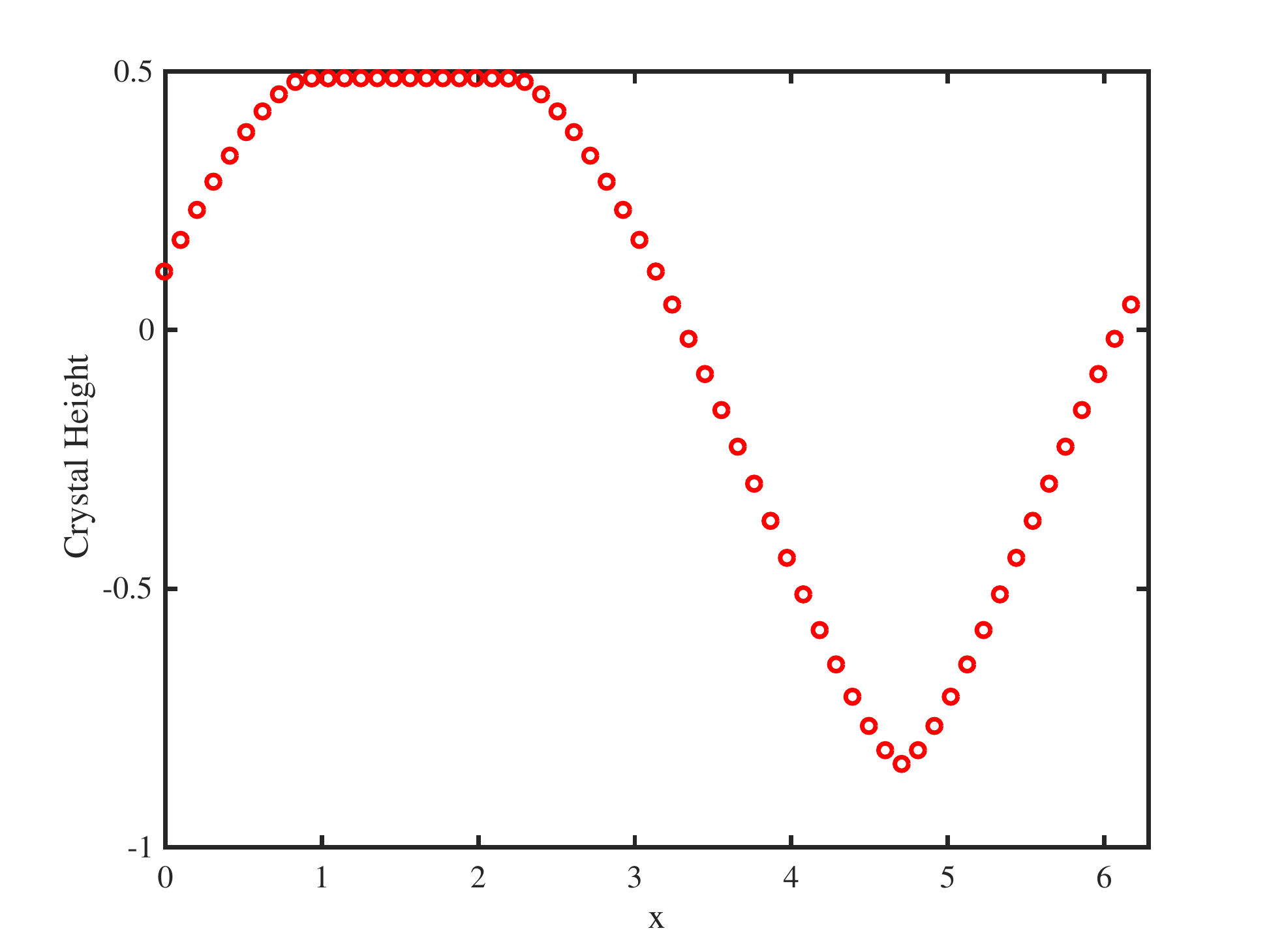}
\caption{Snapshots of surface height evolution by fourth-order, regularized flow \eqref{exppde_visc_reg}.  Left panel: Initial data $h_0 (x) = \sin( x)$. Right panel: Height profile, $h(x,t)$, at $t=T = 10^{-1}$ with $\nu = 10^{-3}$ and $g =3$. }
\label{f:reg}
\end{center}
\end{figure}

In this case, there is no explicit ODE system to predict the dynamics of facets, since the underlying, regularized energy~\eqref{ereg}
does not permit the formation of jumps in height and facets (flat parts of the height profile) for $\nu,\, g > 0$.  Indeed, our numerical scheme does not result in jumps in the height profile in this case, though of course the asymmetry of the exponential model is still manifest in the evolution; 
see~Figure~\ref{f:reg} for a typical evolution of \eqref{exppde_reg} with $g = 3$. For sufficiently small regularization parameter, $\nu$, the numerical solution for $h$  evolves to become quite flat near a maximum. This flat part of the height profile is still considered as a facet. In contrast, the  height profile  near a minimum seems to develop a discontinuity in the slope (see Figure~\ref{f:reg}).
%The height continuity in this framework for any $g >0$ is of course expected %from the physics of step models.

We note that the case with $g>0$ in \eqref{exppde_reg} results in  dynamics similar to those observed in \cite{MW1} with interaction potentials $V(z) = |z|^p$, $p >1$.  These dynamics include a flattening of the surface profile near the maximum of the initial height, and the finite-time formation of a discontinuity in the derivative of the height at the minimum of the initial height profile.  We conjecture that these features are indeed expected in these types of degenerate fourth-order PDEs with exponential mobility.  The reader is referred to~\cite{MW1} for a more detailed discussion of this type of breakdown of regularity in $\partial_x h$ in various settings.

%%%%%%%%%%%%%%%
%%%%%%%%%%%%%%%
\section{Conclusion and discussion} 
\label{sec:Discussion}
%%%%%%%%%%%%%%%
%%%%%%%%%%%%%%%

In this paper, we studied plausible implications of continuum evolution law~\eqref{eq:PDE-evol} which emerges from a mesoscale model for line defects and an atomistic broken-bond model in crystal surface morphological evolution. A noteworthy feature of this PDE is the presence of an exponential, singular mobility which has a significant effect if the Boltzmann energy, $k_BT$, is small compared to the step line tension. \color{black} Because of this feature, the evolution occurs in the framework of a nonlinear, weighted $H^{-1}$ gradient flow. For this evolution, in the absence of elasticity ($g=0$), we constructed a solution for the surface height that explicitly manifested an asymmetry between the dynamics on convex and concave parts of the crystal surface. This asymmetry manifests in the following way. Top facets expand fast, regardless of their initial size; in contrast, bottom facets move only if their size exceeds a certain critical length (see Remark~1).

Our analysis points to several open questions. For example, it is compelling to ask if the predicted facet asymmetry can be observed in one-dimensional periodic gratings in homoepitaxy. \color{black}
So far, we have focused on crystal surfaces in 1+1 dimensions.  However, continuum evolution laws with an exponential mobility in 2+1 dimensions have been derived~\cite{MW1}; in addition, such PDEs are plausibly linked to step flow~\cite{DM_Kohn06}. Therefore, the analysis of the dynamics stemming from such equations in higher dimensions is an interesting topic for future study. A related, pending issue is to understand the effects of (short- or long-range) elasticity on the facet evolution. In this case, the analytical description of facet dynamics is more challenging.

We note that the (non-weighted) $H^{-1}$ total variation flow \eqref{eq:gigatvpde} and the corresponding $L^2$ total variation flow have been studied in some detail by many authors, e.g. \cite{KobayashiGiga99,GigaGiga,GigaKohn,Giga1}.  In the (weighted) $H^{-1}$ setting with an exponential mobility, these studies fall into the more general framework of evolution equations of the form
\[
h_t = \mathcal{L} e^{ \mu[h]}~, \ \ h(x,0) = h_0~,
\]
where $\mathcal L$ is an appropriate second-order differential operator dictated by the dominant kinetic processes in surface diffusion, e.g., $\mathcal{L} = \Delta$ for diffusion-limited kinetics; recall that the step chemical potential, $\mu[h]$, is the variation of the surface free energy.  The analysis of evolutions of this form is still under development, including ODE dynamics for facets, existence of solutions in the total variation norm, and finite relaxation times (otherwise known as extinction times) to reach the equilibrium state in surface morphological relaxation. Similar issues arise in the theory of evolution PDEs of weighted $L^2$ total variation flow. These studies may shed light on the dynamics of various phenomena on crystal surfaces, e.g., the dewetting of thin, solid films~\cite{Chame2013}. 

It is worthwhile noting that the exponential PDE derived from atomistic dynamics in \cite{MW1} with $p=2$ has the form (for $\mathcal L=\Delta$)
\[
\partial_t h = \Delta e^{- \Delta h}.
\]
This PDE is studied in \cite{LX}, where the authors derive weak solutions for a class of functions where $\Delta h$ lives in a measure space.  Extending such derivations and global dynamics to a general family of $4$th-order degenerate PDE models with exponential mobility deserves attention for future research. 

In the present work, we refrained from comparing the facet dynamics predicted by the continuum theory to the underlying microscale dynamics, particularly the motion of steps. An interesting feature in this context is the interaction between steps in the vicinity of a facet. This topic will be the subject of future work.

In a related fashion,  the numerical schemes that we use here are based on straightforward finite-difference discretizations.  Of course, energetic methods such as those for related problems in \cite{KohnV} motivated by algorithms developed in \cite{JKO} would seem viable. However, the presence of the exponential mobility renders these methods much more computationally expensive.  The convergence analysis and development of efficient numerical schemes for evolution equations of form \eqref{eq:PDE-evol} will be valuable for predictions of faceting in crystal surface morphological evolution.

%%%%%%%%%%%%%%%%%%%%%%%%%%%%%%%%%%%%%%%%%%%%%%%%%%
%%%%%%%%%%%%%%%%%%%%%%%%%%%%%%%%%%%%%%%%%%%%%%%%%%

\section*{Acknowledgments}
The authors wish to thank Professors Y. Giga, T.~L. Einstein, R.~V. Kohn and
J.~Q. Weare for valuable discussions. JGL was supported in part by
National Science Foundation (NSF) under award DMS-1514826. JL was
supported in part by NSF under award DMS-1454939.  The research of the
DM was supported by NSF DMS-1412769 at the University of Maryland.
The research of JLM was supported by NSF Grant DMS-1312874 and NSF
CAREER Grant DMS-1352353. This collaboration is made possible thanks
to the NSF grant RNMS-1107444 (KI-Net).

%%%%%%%%%%%%%%%%%%%%%%%%%%%%%%%%%%%%%%%%%%%%%%%%%%
%%%%%%%%%%%%%%%%%%%%%%%%%%%%%%%%%%%%%%%%%%%%%%%%%%

%\appendix

%%%%%%%%%%%%%%%%%%%%%%%%%%%%%%%%%%%%%%%%%%%%%%%%%%
%%%%%%%%%%%%%%%%%%%%%%%%%%%%%%%%%%%%%%%%%%%%%%%%%%

%%%%%%%%%%%%%%%%%%%%%%%%%%%%%%%%%%%%%%%%%%%%%%%%%%%%%%%
%\section{} 

%%%%%%%%%%%%%%%%%%%%%%%%%%%%%%%%%%%%%%%%%%%%%%%%%%%%%%%


\begin{thebibliography}{99}


\bibitem{AlHajjShehadehKohnWeare:2011}
H. Al Hajj Shehadeh, R.~V. Kohn, J. Weare,  
The evolution of a crystal surface: analysis of a one-dimensional step train connecting two facets in the ADL regime, Physica D 240 (2011) 1771--1784.

\bibitem{AGS}
L. Ambrosio, N. Gigli, and G. Savar\'e,
Gradient flows: in metric spaces and in the space of probability measures. Springer Science \& Business Media (2008).

\bibitem{BonitoNQM_09}
 A. Bonito, R.~H. Nochetto, J. Quah, D. Margetis,
 Self-organization of decaying surface corrugations: a numerical study,
 Phys.\ Rev.\ E 79 (2009) 050601(R).
 
 \bibitem{BonzelPreuss95}H.~P. Bonzel, E. Preuss, Morphology of periodic surface profiles below the roughening temperature: aspects of continuum theory, 
Surf.\ Sci.\ 336 (1995) 209--224.

\bibitem{BCF51}W.~K. Burton, N. Cabrera, F.~C. Frank, The growth of crystals and the equilibrium structure of their surfaces, Philos.\ Trans.\ R.\ Soc.\ Lond.\ Ser.\ A 243 (1951) 299--358.

\bibitem{Chame2013}A. Chame, O. Pierre-Louis,
Modeling dewetting of ultra-thin solid films,
Comptes Rendus Physique 14 (2013) 553--563.


\bibitem{Chan04}W.-L. Chan, A. Ramasubramaniam, V.~B. Shenoy, E. Chason,  Relaxation kinetics of nano-ripples on Cu(001) surface,
Phys.\ Rev.\ B 70 (2004) 245403.

\bibitem{Freund-book}L.~B. Freund, S. Suresh, Thin Film Materials: Stress, Defect Formation and Surface Evolution, Cambridge University Press, Cambridge, UK, 2009.

\bibitem{FLL}
I. Fonseca, G. Leoni, Y.~Y. Lu, Regularity in time for weak solutions of a continuum model for epitaxial growth with elasticity on vicinal surfaces, Commun.\ Partial Diff.\ Equations 40 (2015) 1942--1957.

\bibitem{Funaki}T. Funaki, Stochastic Interface Models,
  Lectures on Probability Theory and Statistics, Lecture Notes in
    Mathematics 1869, Springer, Berlin, 2005, pp.\ 103--274.
  
  \bibitem{funakispohn}T. Funaki, H. Spohn, Motion by mean curvature for the Ginzburg-Landau $\nabla \phi$ interface model,
  Commun.\ Math.\ Phys.\ 185 (1997) 1--36.


\bibitem{GigaGiga}M.~H. Giga, Y. Giga, Very singular diffusion equations: second and 
fourth order problems, Japan J.\ Indust.\ Appl.\ Math.\ 27 (2010) 323--345.

\bibitem{GigaKohn}Y. Giga, R.~V. Kohn, 
Scale-invariant extinction time estimates for some singular diffusion equations,
Discr.\ Cont.\ Dyn.\ Sys.\  A 30 (2011) 509--535. 

\bibitem{Giga1}Y. Giga, H. Kuroda, H. Matsuoka, Fourth-order total variation flow with Dirichlet condition:  characterization of evolution and extinction time estimates,
Adv.\ Math.\ Sci.\  App.\  24 (2014) 499--534.

\bibitem{GruberMullins67}E.~E. Gruber, W.~W. Mullins, On the theory of anisotropy of crystalline surface tension,
J.\ Phys.\ Chem.\ Solids 28 (1967) 875--887.

\bibitem{IhleMisbahP-L98} T. Ihle, C. Misbah, O. Pierre-Louis. Equilibrium step dynamics on vicinal surfaces revisited, Physical Review B 58.4 (1998) 2289.

\bibitem{IsraeliKandel99}
N. Israeli, D. Kandel, 
Profile of a decaying crystalline cone,
Phys.\ Rev.\ B 60 (1999) 5946--5962.

\bibitem{IsraeliKandel00}N. Israeli, D. Kandel, Decay of one-dimensional surface modulations, Phys.\ Rev.\ B 62 (2000) 13707--13717.


\bibitem{IsraeliJeongKandelWeeks:2000:stepscaling} 
N. Israeli, H.-C. Jeong, D. Kandel, J.~D. Weeks, 
Dynamics and scaling of one-dimensional surface structures,
Phys.\ Rev.\ B 61 (2000)  5698--5706.


\bibitem{JeongWilliams99}H.-C. Jeong, E.~D. Williams, Steps on surfaces: experiment and theory, Surf.\ Sci.\ Reports 34 (1999) 171--294.

\bibitem{JKO}R. Jordan, D. Kinderlehrer, F. Otto, 
The variational formulation of the Fokker-Planck equation,
SIAM J.\ Math.\ Anal.\ 29 (1998) 1--17. 

\bibitem{Kashima}Y. Kashima, A subdifferential formulation of fourth order singular diffusion equations, Adv.\ Math.\ Sci.\ Appl.\ 14 (2004) 49--74.

\bibitem{KobayashiGiga99}R. Kobayashi, Y. Giga, Equations with singular diffusivity, J.\ Stat.\ Phys.\ 95 (1999) 1187--1220.


\bibitem{KohnV}R.~V. Kohn, E. Versieux, Numerical analysis of a steepest-descent PDE model for surface relaxation below the roughening temperature, SIAM J.\ Num.\ Anal.\ 48 (2010) 1781--1800.

\bibitem{Sethna96}B. Krishnamachari, J. McLean, B. Cooper, J. Sethna,  Gibbs-Thomson formula for
small island sizes: corrections for high vapor densities, Phys.\ Rev.\ B 54 (1996) 8899--8907.

\bibitem{KDM}J. Krug, H.~T. Dobbs, S. Majaniemi, Adatom
    mobility for the solid-on-solid model, Z.\ Phys.\ B 97 (1995) 281--291.
    
\bibitem{Kukta02-I}R.~V. Kukta, K. Bhattacharya, A micromechanical model of surface steps, J.\ Mech.\ Phys.\ Solids 50 (2002) 615--649.

\bibitem{Kukta02-II}R.~V. Kukta, A. Peralta, D. Kouris, Elastic interaction of surface steps: effect of
atomic-scale roughness, Phys.\ Rev.\ Lett.\ 88 (2002) 186102.
    
\bibitem{LX}J.-G. Liu, X. Xu, 
Existence theorems for a multidimensional crystal surface model,
SIAM J.\ Math.\ Anal.\ 48 (2016) 3667--3687. 
    
\bibitem{LLM}J. Lu, J.-G. Liu, D. Margetis, Emergence of step flow from an atomistic scheme of epitaxial growth in $1+1$ dimensions, Phys.\ Rev.\ E 91 (2015) 032403.

\bibitem{MarchenkoP80}V.~I. Marchenko, A.~Ya. Parshin, Elastic properties of crystal surfaces, Soviet Phys.\ JETP 52 (1980) 129--131.
    
    \bibitem{MFAS06}D. Margetis, P.-W. Fok, M.~J. Aziz, H.~A. Stone, Continuum theory of nanostructure decay via a microscale condition,
Phys.\ Rev.\ Lett.\ 97 (2006) 096102.


\bibitem{DM_Kohn06}D. Margetis, R.~V. Kohn, Continuum relaxation of interacting steps on crystal surfaces in 2+1 dimensions, Multiscale Model.\ Simul.\ 5 (2006) 729--758. 

\bibitem{MW1}J.~L. Marzuola, J. Weare, The relaxation of a family of broken bond crystal surface models,
 Phys.\ Rev.\ E 88 (2013) 032403.




\bibitem{Misbah10}C. Misbah, O. Pierre-Louis, Y. Saito, Crystal surfaces in and out of equilibrium: a modern view, Rev.\ Mod.\ Phys.\ 82 (2010) 981--1040. 

\bibitem{Srolovitz94}R. Najafabadi, D.~J. Srolovitz, 
Elastic step interactions on vicinal surfaces of fcc metals,
Surface Sci.\ 317 (1994) 221--234.

\bibitem{Nishikawa}T. Nishikawa, Hydrodynamic limit for the Ginzburg-Landau $\nabla \phi$ interface model with a conservation law, 
J.\ Math.\ Sci.\ Univ.\ Tokyo 9 (2002) 481--519.

\bibitem{Odisharia_06}I.~V. Odisharia, Simulation and Analysis of the Relaxation of a Crystalline Surface, Ph.~D. Thesis, Courant Institute, New York University, 2006.

\bibitem{OZ90} 
 M. Ozdemir, A. Zangwill,  Morphological equilibration of a corrugated crystalline surface,
Phys.\ Rev.\ B 42 (1990) 5013--5024. 


\bibitem{PimpinelliVillain98}A. Pimpinelli, J. Villain, Physics of Crystal Growth, Cambridge University Press, Cambridge, UK, 1999.

\bibitem{RV88}
 A. Rettori, J. Villain, Flattening of grooves on a crystal surface: a method of investigation of surface roughness, J.\ Phys.\ 49 (1988) 257--267.


\bibitem{Rowlinson}J.~S. Rowlinson, B. Widom, Molecular Theory of Capillarity, Clarendon Press, Oxford, 1982.

\bibitem{Shenoy02}V.~B. Shenoy, L.~B. Freund, A continuum description of the energetics and evolution
of stepped surfaces in strained nanostructures, J.\ Mech.\ Phys.\ Solids 50 (2002) 1817--1841.

\bibitem{Shenoy04}V.~B. Shenoy, A. Ramasubramaniam, H. Ramanarayan, D.~T. Tambe, W.-L. Chan, E. Chason, Influence of step-edge barriers on the morphological relaxation of nanoscale ripples on crystal surfaces, Phys.\ Rev.\ Lett.\ 92 (2004) 256101.

\bibitem{Spohn93}H. Spohn, Surface dynamics below the roughening transition, 
J.\ Phys.\ I (France) 3 (1993) 69--81.

\bibitem{Tanaka97}S. Tanaka, C.~C. Umbach, J.~M. Blakely, Atomic step distributions on annealed periodic Si(001) gratings, J.\ Vac.\ Sci.\ Technol.\ A 15 (1997) 1345--1350.


\bibitem{Weeks79}J.~D. Weeks, G.~H. Gilmer, Dynamics of crystal growth, 
Adv.\ Chem.\ Phys.\ 40 (1979) 157--228.


\bibitem{XX}H. Xu, Y. Xiang, Derivation of a continuum model for the long-range elastic interaction on stepped epitaxial surfaces in 
$2+1$ dimensions, SIAM J.\ Appl.\ Math.\ 69 (2009) 1393--1414.  

\bibitem{Zangwill91}A. Zangwill, C.~N. Luse, D.~D. Vvedensky, M.~R. Wilby, Equations of motion for epitaxial growth , Surf. Sci. Let..\ 274 (1991) 529--534.

\bibitem{Zangwill92}A. Zangwill, C.~N. Luse, D.~D. Vvedensky, M.~R. Wilby, Epitaxial growth and recovery: an analytical approach, Mater.\ Res.\ Soc.\ Symp.\ Proc.\ 237 (1992) 189--198.







 
 
\end{thebibliography}
\end{document}